\documentstyle[11pt]{article}
\newcommand{\dddot}[1]{\stackrel{...}{#1}}

\newcommand{\be}{\begin{eqnarray*}}
\newcommand{\ee}{\end{eqnarray*}}
\newcommand{\bd}{\begin{description}}
\newcommand{\ed}{\end{description}}
\newcommand{\bi}{\begin{itemize}}
\newcommand{\ei}{\end{itemize}}
\newcommand{\bc}{\begin{center}}
\newcommand{\ec}{\end{center}}

\newcommand{\nn}{\nonumber}

\input amssym.def

\begin{document}

\title{Anisotropic Homogeneous Cosmologies in the Post-Newtonian
Approximation. }
\author{Tamath Rainsford\\
Department of Physics and Mathematical Physics, \\
University of Adelaide, South Australia 5005, Australia}
\date{}
\maketitle

\begin{abstract}
\noindent 
In this paper we explore how far the post-Newtonian theory, \cite{szek-rain}
goes in overcoming the difficulties associated with anisotropic homogeneous
cosmologies in the Newtonian approximation. It will be shown that, unlike in
the Newtonian case, the cosmological equations of the post-Newtonian
approximation are much more in the spirit of general relativity with regard
to the nine Bianchi types and issues of singularities.

The situations of vanishing rotation and vanishing shear are treated
separately. The homogeneous Bianchi I model is considered as an example of a
rotation-free cosmology with anisotropy. It is found in the Newtonian
approximation that there are arbitrary functions that need to be given for
all time if the initial value problem is to be well-posed, while in the
post-Newtonian case there is no such need. For the general case of a perfect
fluid only the post-Newtonian theory can satisfactorily describe the effects
of pressure. This is in accordance with findings in \cite{rain} where the
post-Newtonian approximation was applied to homogeneous cosmologies.

For a shear-free anisotropic homogeneous cosmology the
Newtonian theory of Heckmann and Sch\"ucking, \cite{heck-shuck55} is
explored. Comparisons with
its relativistic and  post-Newtonian counterparts are made. In the Newtonian
theory solutions exist to which there are no analogues in general
relativity. The post-Newtonian approximation may provide a way out.
\end{abstract}

\newpage

\section{Introduction}

Due to its simplicity, the Newtonian approximation of cosmology is
preferable to general relativity and is typically used wherever
possible. However, in a recent paper \cite{szek-rain}, it is shown that the
Newtonian approximation has some difficulties. To begin with the Newtonian
approximation is incomplete in the sense that the Bianchi identities cannot
be obtained from the field equations. It is generally accepted that
Newtonian theory is a specialization of the linearized approximation of
general relativity cut off at the $c^{-2}$-level. However, information is
lost in this cut-off procedure, and it is shown that cutting off at the
$c^{-4}$-level results in a theory in which the Bianchi identities are
obtainable from the field equations, \cite{szek-rain}. Moreover, the
Newtonian theory has no initial value problem when applied to cosmology
because there are no boundary conditions. Therefore, the Poisson equation has
no unique solution. The post-Newtonian theory, on the other hand, is a
closed, consistent and well-posed theory. The post-Newtonian approximation
used here is obtained by expanding up to order $c^{-6}$ and reformulating the
linearized field equations as wavelike equations. A specialization of the
resulting equations leads to a formally well-posed $c^{-4}$-theory. This 
$c^{-4}$-approximation is the first order approximation to give us
consistency and formal well-posedness and, as we shall see, suffices as a
``Newtonian'' approximation to cosmology. Going to higher order simply gives
increasingly more accurate approximations of general relativity. 

In \cite{rain} it was shown that the post-Newtonian approximation, in the
context of the $k = 0$ Friedmann-Robertson-Walker cosmology (FRW), is able
to fully reproduce the results of its general relativistic counterpart,
whilst the Newtonian theory is not. The Newtonian approximation of cosmology
does not allow the pressure to enter into the dynamics. Hence, changing the
equation of state, does not result in different solutions for the density and
expansion. In fact, the only solution that agrees with general relativity
is that for the matter dominated universe. On the other hand, in the
post-Newtonian approximation pressure becomes dynamic
through an extra function of time, $A(t)$. Hence, the post-Newtonian theory is
able to produce differing solutions for alternative equations of state in
accordance with
general relativity. In the case where $A(t)$ goes to zero the
pressure is forced to vanish and the higher order terms of the
post-Newtonian approximation vanish as well, giving exactly the same solution
for dust as the Newtonian theory does. Thus, Newtonian theory for
homogeneous and isotropic cosmology should only be used for the special case
of dust, where the two theories, the Newtonian and post-Newtonian, coincide.

In this paper we explore how far the post-Newtonian approximation goes in
overcoming the difficulties of the Newtonian theory when applied to the more
general case of anisotropic homogeneous cosmology. There is a great deal of
theoretical and observational evidence to support anisotropy in
the universe, \cite{mac}.  Any theory which attempts to approximate general
relativity should yield similar results to the fully relativistic theory. The
FRW models are isotropic and homogeneous. Although they fit most of the
observed features of the universe they do not fit all. Different geometries
may be worth considering at earlier times, for example, near the initial
singularity. There might exist a general rotation of matter in the universe
of comparable magnitude to the general expansion that we can not
detect. Shear might provide a deviation away from isotropy since
extra-galactic objects might be observed by the galactic equator
\cite{heck-shuck59}. All homogeneous cosmologies fall into two classes: the
Bianchi models, which are those for which the isometry group admits a
3-dimensional simply transitive subgroup, and those for which the isometry
group is neither simply transitive, nor admits a simply transitive subgroup --
the Kantowski-Sachs models. There exist nine Bianchi types and,
correspondingly, nine Bianchi cosmologies, each class has subclasses with
extra symmetries. The Bianchi types are in general anisotropic, so they do
not have all spatial directions at a point being equivalent. Instead, there
are now accelerations, and anisotropic observers will no longer be orthogonal
to the surfaces of homogeneity. Not all Bianchi types are suitable for
describing the universe. They were first considered as cosmological models by
Taub, \cite{taub} then shortly followed by Raychaudhuri, \cite{ray55} and
Heckmann and Sch\"ucking, \cite{heck-shuck62}. These models contain the FRW models as
a special case. Thus, they are considered the simplest and the most likely
cosmological candidates, in particular Bianchi I, Bianchi IX and the LRS
types $7_o$ and $7_h$.

Bianchi I has a different cosmic scale factor for each direction and only
becomes isotropic in the case where the scale factors all become equal. In
this case the Bianchi I model becomes the $k = 0$ FRW
cosmology. For nonzero shear these rotation-free anisotropic models
are called the Heckmann-Sch\"ucking solutions,
\cite{heck-shuck55}.\footnote{Not to be confused with the Newtonian
Heckmann-Sch\"ucking cosmologies.} In the vacuum case they become the Kasner
solution.

Unfortunately, there are solutions in the case of anisotropy to which there
are no general relativistic analogues. Theorems of Ellis, \cite{ellisthrm}
tell us that shear-free perfect fluids have either vanishing expansion or
vanishing rotation. The case of vanishing rotation just yields the FRW model
which is explored in \cite{rain}. The case of vanishing expansion, however,
is more problematic since in Newtonian cosmology it allows for
singularity-free solutions. This in in contradiction with theorems of Hawking
which state that solutions of general relativity must have singularities. The
post-Newtonian approximation provides a way out of this difficulty.

We begin in section 2.1 with a review of homogeneous Newtonian cosmology. 
Then in 2.2 we consider post-Newtonian approximation to homogeneous
anisotropic cosmology. We find that the Newtonian theory is in general
under-determined and unable to fully reproduce the nine Bianchi types of
general relativity. But the post-Newtonian equations produce a set of
cosmological equations which are more in the spirit of the Bianchi
identities.

In section 3 we consider the case where the rotation vanishes
separately. Bianchi I is taken as an example of a rotation-free anisotropic
and homogeneous cosmology, and explored in the Newtonian and post-Newtonian
contexts. In section 4 we consider the shear-free case. We end in section
5 with a conclusion.

\newpage

\section{Newtonian Theory and the Post-Newtonian Approximation in Homogeneous
Anisotropic Cosmologies}

\subsection{Newtonian Homogeneous Anisotropic Cosmology}

In the following we will consider the Newtonian cosmological model of
Heckmann and Sch\"ucking, \cite{heck-shuck55}. For a homogeneous cosmology the
density, $\rho$, and the pressure, $p$, are functions of time only. The
velocity field, $v_i$, is the same relative to all observers and such that
$v_i = V_{ij}(t)X_j$, \cite{m-r,heck-shuck55,szek-rank}. The theory is then
described by the Poisson, continuity and Euler equations, which are given by:
\begin{eqnarray}
\phi_{,ii} & = & 4\pi G \rho, \label{c7Poisson} \\
\dot{\rho} + \rho v_{i,i} & = & 0, \label{c7*} \\
\dot{v_i} + \phi_{,i} & = & 0, \label{c7**}
\end{eqnarray}
where $G$ is the Newtonian gravitational constant. Here $\dot{\bullet}$
refers to the convective derivative, $\dot{\bullet} =
\frac{\partial}{\partial t}\bullet + v_i \bullet_{,i} $

We find from (\ref{c7**}) that the Newtonian potential $\phi$, up to a
constant, takes the form $\phi =
a_{ij}(t)X_iX_j$. Thus equation (\ref{c7Poisson}) simplifies to

\begin{eqnarray}
a_{ii} = 4\pi G\rho. \label{c7poisson}
\end{eqnarray}
We make the following decomposition
\begin{eqnarray}
V_{ij} = \frac{1}{3} \theta \delta_{ij} + \sigma_{ij} + \omega_{ij},
\label{c7decomposition}
\end{eqnarray}
with
\begin{eqnarray}
\theta & = & V_{ii}, \nn \\
\sigma_{ij} & =  & \frac{1}{2} (V_{ij} + V_{ji}) 
        - \frac{1}{3} \theta \delta_{ij}, \nn \\
\omega_{ij} & = & \epsilon_{jik} \omega_k = \frac{1}{2}(V_{ij} - V_{ji}), \nn
\end{eqnarray}
where the trace, $\theta$ is the expansion, the trace-free
symmetric part $\sigma_{ij}$ is the shear and the anti-symmetric part $\omega_{ij}$
is the rotation. Inserting these expressions for $V_{ij}$, equations
(\ref{c7*}) and (\ref{c7**}) become 
\begin{eqnarray}
\dot{\rho} + \rho \theta = 0, \label{c7cont}
\end{eqnarray}
and
\begin{eqnarray}
\dot{\theta} & = & 4\pi G\rho  - \frac{1}{3} \theta^2 + 2\omega^2 - 2 \sigma^2,
\label{c7teta} \\
\dot{\omega}_i & = & - \frac{2}{3} \theta \omega_i + \sigma_{ij}\omega_j,
\label{c7w} \\ 
\dot{\sigma}_{ij} & = & - \frac{2}{3} \theta \sigma_{ij} - \sigma_{ik}
\sigma_{kj} - \omega_{ik} \omega_{kj} + \frac{2}{3}(\sigma^2 - \omega^2) \delta_{ij} -
a_{ij} + \frac{1}{3} a_{kk} \delta_{ij}, \label{c7sigma}
\end{eqnarray}
where $\sigma^2 = \frac{1}{2}\sigma_{ij}\sigma_{ij}$ and $\omega^2 =
\frac{1}{2}\omega_{ij}\omega_{ij}=\omega_i\omega_i$.

Equations (\ref{c7poisson}) and (\ref{c7cont})--(\ref{c7sigma}), along with a
barotropic equation of state relating the density and pressure, form the
Newtonian approximation to cosmology. The structure of these equations is as
follows: There are eleven equations, one each arising from the Poisson,
continuity and the trace of the Euler equation, five from the symmetric
trace-free part and a further three from the
antisymmetric part. However, there are 16 unknowns: $\theta$, $\rho$,
$\omega_i$, $\sigma_{ij}$ and $a_{ij}$.  Thus, the system is under-determined
and it is not possible to solve uniquely for all the unknowns.

Providing five arbitrary functions of time, for instance by specifying the
shear $\sigma_{ij}$(t), for all time, we are able to solve for the five
functions $\theta$, $\rho$ and $\omega$, the Newtonian potential $\phi =
a_{ij}(t)X_i X_j$ and the pressure $p$ through the barotropic equation of
state. Then, for a set of initial data $(\omega_i(x_i,0),\rho(t_0), \phi(t_0))$
the set of homogeneous Newtonian cosmologies may be found since (\ref{c7cont}),
(\ref{c7teta}) and (\ref{c7w}) can be uniquely solved for $\theta$, $\omega_{ij}$ and
$\rho$, and, $a_{ii}$ can be determined from (\ref{c7poisson}) and $a_{ij}$
from (\ref{c7sigma}).

This is very different to the fully general relativistic case where there are
nine possible symmetry groups.  In general relativity the system of equations
which describe cosmology has a well-posed Cauchy problem. In the Newtonian
case, however, the system is under-determined - at least five functions of the
set $\theta$, $\omega_i$ and $\sigma_{ij}$ must be specified for all time.
Thus, the Newtonian theory is not able to fully reproduce the full Bianchi
types of general relativity in general. Arbitrary shear has no analogy in
general relativity. However, once five (arbitrary) constraints have been
imposed, the theory has an initial value formulation within the constraint of
being homogeneous and isotropic and within the strange set up of allowing a
given evolution of the shear.

Let us explore how far the post-Newtonian approximation \cite{szek-rain} goes
in overcoming this difficulty. The post-Newtonian theory is known to be
well-posed in the general case of anisotropy and inhomogeneity
\cite{szek-rain}. The extra potentials of the post-Newtonian theory act as
integrability conditions for the Newtonian potential $\phi$. We will see how
these extra potentials help us in making the system determined in general.

\subsection{Post-Newtonian Approximations of Anisotropic Homogeneous
Cosmologies}

Following a scheme similar to that of Weinberg \cite{wein}, we adopt units in
which the typical velocity has magnitude 1, i.e.\ $\beta \approx
\frac{v}{c}$, and assume a one parameter family of metrics
$g_{\mu\nu}(x^{\lambda},c)$ for which there is a system of coordinates $(x^0,
x^i)$ in which the components of the metric have the following asymptotic
behaviour as $c \longrightarrow \infty$:
\begin{eqnarray}
g_{00} & = & -1 -2 \phi c^{-2} - 2 \alpha c^{-4} - 2 \alpha' c^{-6} - 2
\alpha''c^{-8}..... \;, \nonumber \\
g_{0i} & = & \zeta_i c^{-3} + \zeta_i' c^{-5} + \zeta_i'' c^{-7}.....\;, \nn \\
g_{ij} & = & \delta_{ij} - 2 \phi \delta_{ij} c^{-2} + \alpha_{ij} c^{-4}
                + \alpha_{ij}' c^{-6} + \alpha_{ij}''
c^{-8}.....\;.\label{c7pnform}
\end{eqnarray}

The usual Newtonian theory is obtained as the ${\cal O}(c^{-2})$ limit of
(\ref{c7pnform}), while the Newtonian approximation is the ${\cal
O}(c^{-4})$ limit. Reformulating the field equations of the ${\cal
O}(c^{-6})$ limit as wavelike equations defines the post-Newtonian theory
used here (see \cite{szek-rain}).

After applying the harmonic gauge conditions
\begin{eqnarray}
\dot{\phi} & = & - \frac{1}{4}\zeta_{i,i},\label{c7hg2}\\
\dot{\zeta_i} & = & \phi_{ij,j}, \label{c7hg1}
\end{eqnarray}
the field equations for the post-Newtonian theory are
\begin{eqnarray}
\phi_{,kk} & = & 4\pi G \rho + \frac{1}{4c^2}\left(- \phi_{jk,jk} - J\right),
\label{c7C1}
\\
\zeta_{i,kk} & = & 16 \pi G\rho v_i  + \frac{1}{c^2} \left(
\dot{\phi}_{ij,j} - K_i \right),  \label{c7C2i} \\
\ddot\phi_{ij} - c^2\phi_{ij,kk} & = &  K_{ij} + c^2\left[16\pi G
(\rho v_i v_j + \delta_{ij} p) - J_{ij} \right],
\label{c7N3ij}
\end{eqnarray}
where the $\alpha$ and $\alpha_{ij}$ of (\ref{c7pnform}) are such that
\begin{eqnarray}
\phi_{ij} = \alpha_{ij} - 2 \delta_{ij} \alpha, \nn
\end{eqnarray}
with
\begin{eqnarray}
J & \equiv & 6\phi_{,i}\phi_{,i} - 16 \pi G (\rho v^2 + 4\rho\phi), \nn
\\
K_i & \equiv &  3\zeta_{j,j}\phi_{,i} +
2\zeta_j\phi_{,ij} - 2\phi_{,j}\zeta_{j,i}\nonumber \\
& & \quad - 16\pi G\left[v_i p + \rho v_iv^2 -
\frac{1}{2}\rho \zeta_i\right], \nn\\
J_{ij} & \equiv & 8\phi\phi_{,ij} + 4\phi_{,i}\phi_{,j} - \delta_{ij}
(6\phi_{,k}\phi_{,k} + 32\pi G\rho\phi),\nn \\
K_{ij} & \equiv &
  -\frac{1}{2}\left(
        \zeta_i\zeta_{k,kj} + \zeta_j\zeta_{ k,ki}
   \right)
  - \zeta_k\left(
        \zeta_{i,jk} + \zeta_{j,ik}
   \right) + 2\zeta_k\zeta_{k,ij} + \zeta_{k,i}\zeta_{k,j}
  + \zeta_{i,k}\zeta_{j,k}\nonumber \\
&&
  - 2\phi_{,k}\left(
        \phi_{ki,j} + \phi_{kj,i} - 2\phi_{ij,k}
   \right) - 16\phi\phi_{,i}\phi_{,j} + \phi_{,i}\phi_{kk,j} \nonumber\\
&&
  + \phi_{,j}\phi_{kk,i} - 2\phi\left(
        \phi_{ki,jk} + \phi_{kj,ik} - \phi_{ij,kk} - \phi_{kk,ij}
   \right) - 2\phi_{ki}\phi_{,jk} \nonumber\\
&&
  - 2\phi_{kj}\phi_{,ik} + 2\phi_{,ij}\phi_{kk} - \delta_{ij}\left(
        \frac{1}{2}\zeta_{k,m}\zeta_{k,m} + \frac{1}{2}\zeta_{m,k}\zeta_{k,m}
        + \frac{1}{2}(\zeta_{k,k})^2
   \right.\nonumber\\
&&
  - \zeta_k\zeta_{m,mk} - 4\phi_{,k}\phi_{km,m} + 4\phi_{,k}\phi_{mm,k}
  - 12\phi\phi_{,k}\phi_{,k} \nonumber\\
&&
        - \phi\left(
                2\phi_{km,mk} - 2\phi_{mm,kk}
          \right)\bigg)  \nn\\
&&
  + 8\pi G\Bigg[
        2pv_iv_j
    + 2\rho(2\phi + v^2)v_iv_j + \rho\phi_{ij} \nn\\
&&
        + \delta_{ij}\left(
           2\rho \phi v^2 - \frac{1}{2}\phi p
           + \frac{3}{4}\phi_{,k}\rho_{,k} + \frac{1}{2}\rho \phi_{,kk}
        \right)
  \Bigg]. \nn
\end{eqnarray}

This system forms a closed set which is consistent because the
Bianchi identities are obtainable from the field equations:
\begin{eqnarray}
&& \dot\rho \left(1 + \frac{v^2 - 4\phi}{c^2}\right)
        + (\rho v_j)_{,j}\left(1 + \frac{v^2}{c^2}\right) 
        + \frac{1}{c^2} \left[ \rho\left(2v_j\dot v_j + 2v_jv_k v_{k,j}
        +\frac{1}{2}\zeta_{j,j}\right)\right. \nonumber\\
& & \left. - \frac{1}{2}\rho_{,j}\zeta_j + (v_jP)_{,j} 
        + \frac{1}{16 \pi G}(2\phi_{,i}\zeta_{i,jj} - 2\zeta_i\phi_{,jii}
        - 3\zeta_{i,i}\phi_{,jj}) \right] = 0
\label{c7B1}
\end{eqnarray}
and
\begin{eqnarray}
&& \rho(\dot{v}_i + v_{i,j}v_j + \phi_{,i}) + P_{,i} 
        = \frac{1}{16\pi G c^2} \left[ -(\dot{J} + K_{j,j})v_i + \right. 
        \nonumber \\
& &  \left. \dot{K}_i - K_{ij,j} - 2\phi(J_{,i} +\phi_{jk,kij}) 
        - \phi_{,i}(J + \phi_{jk,jk}) \right]. \label{c7B2i}
\end{eqnarray}

Homogeneity can be provided by the following ans\"atze:
\begin{eqnarray}
\phi & = & a_{ij}(t) X_{ij} + a(t), \nn \\
\zeta_i & = & b_{ijkl}(t) X_{jkl} + b_{ij}(t)X_j, \label{c7ansatze} \\
\phi_{ij} & = & c_{ijklmn}(t)X_{klmn} + c_{ijkl}(t)X_{kl} + c_{ij}(t), \nn
\end{eqnarray}

\noindent where $X_{ij} = X_i X_j$, $X_{ijk} = X_i X_j X_k$ etc. These are
more general ans\"atze than in \cite{rain} where isotropy has been ensured
through similar constraints. As before we assume that the density and
pressure are functions of time only, and are related through a barotropic
equation of state. The velocity field decomposes as shown before in
equation~(\ref{c7decomposition}).

Substituting the ans\"atze (\ref{c7ansatze}) into the field equations for the
post-Newtonian theory, Eqs. (\ref{c7C1} to \ref{c7N3ij}), and comparing
expressions of different powers in $X_i$, yields the set of equations
for the anisotropic homogeneous post-Newtonian approximation:
\begin{eqnarray}
&& a_{kk} = 2\pi G \rho + c^{-2}\left[ - \frac{1}{4} c_{klkl} + 8 \pi G \rho
a \right], \label{c7FE1} \\[1cm]
&& 3\left(2a_{ik}a_{jk} + c_{klklij}\right) - 4\pi G \rho
\left( 4 a_{ij} + V_{ki}V_{kj} \right) = 0, \label{c7FE2} \\[1cm]
&& 3b_{ikkj} = 8\pi G \rho V_{ij} + c^{-2}\left[\dot{c}_{ikkj} -
3b_{kk}a_{ij} - 2a_{ik}b_{kj} + 2a_{jk}b_{ki} + 8\pi G\left(V_{ij}p -
\frac{1}{2}\rho b_{ij}\right)\right], \label{c7FE3} \\[1cm] 
&& 2\dot{c}_{immjkl} = 
    9a_{i\;\fbox{$\scriptstyle j$}} b_{\fbox{$\scriptstyle kl$}\;mm} 
    - 6a_{m\;\fbox{$\scriptstyle j$}} b_{mi\;\fbox{$\scriptstyle kl$}} 
        + 2a_{im}b_{mjkl}
        - 4\pi G \rho \left(- b_{ijkl} 
        + 4 V_{i\;\fbox{$\scriptstyle j$}} V_{m\;\fbox{$\scriptstyle k$}} 
        V_{m\;\fbox{$\scriptstyle l$}} \right),\nn\\
&&
\label{c7FE4} \\[1cm]
&& 2c_{ijkk} - 16 aa_{ij} + 16\pi G (2\rho a + p)\delta_{ij} +
c^{-2}\left[-\ddot{c}_{ij} + 2b_{ik}b_{jk} + 4 a_{ij}c_{kk} - 4 a_{ik}c_{jk} -
4a_{jk}c_{ik} \right. \nn \\
&& \left. - \delta_{ij} \left(\frac{1}{2}b_{kk}b_{ll} + b_{kl}b_{kl}\right) 
+ 8\pi G \left(\rho c_{ij} + \delta_{ij}\left(- \frac{1}{2}  a p 
+ \frac{1}{2}\rho c_{kk}\right)\right)\right] = 0, \label{c7FE5} \\[1cm]
&& 12c_{ijmnkk} - 8\left(2a_{ij}a_{mn} + a_{im}a_{jn} + a_{in}a_{jm}
        - 3 \delta_{ij} a_{mk}a_{nk}\right) 
        + 32\pi G \rho \delta_{ij}a_{mn} \nn\\
&&
\qquad + 8 \pi G \rho\left(V_{im}V_{jn} + V_{in}V_{jm}\right)\nn\\
&&        
+ c^{-2}\Bigg[- \ddot{c}_{ijmn} + 16 a \left( 3 \delta_{ij} a_{km}a_{kn}
        - 2 a_{im}a_{jn} - 2 a_{in}a_{jm} \right) \nn\\ 
& &
\qquad + \frac{3}{2}\bigg( 2 b_{in}b_{kkjm} + 2 b_{jn}b_{kkim} 
        + 2 b_{im}b_{kkjn} + 2 b_{jm}b_{kkin}  - b_{ik} b_{jkmn} 
        - b_{jk} b_{ikmn}\bigg) 
        \nn\\
& &
\qquad + 2\left(a_{in} c_{kkjm} + a_{jn}c_{kkim} + a_{im} c_{kkjn} 
        + a_{jm}c_{kkin}\right) - 4\left( a_{ik}c_{kjmn} + a_{jk}c_{ikmn} 
        - a_{ij}c_{kkmn} \right) \nn\\
& &
\qquad  + 4\pi G\bigg( 2 \rho c_{ijmn} + pV_{im}V_{nj} + pV_{in}V_{mj}
        + pV_{jm}V_{ni} +  pV_{jn}V_{mi} \nn\\
& & 
\qquad\qquad  + 4\rho a \left(V_{im}V_{jn} + V_{jm}V_{in}
        + \delta_{ij} V_{lm}V_{ln} \right) 
        - pa \delta_{ij} a_{mn} + \delta_{ij} c_{mnkk}\bigg)\Bigg] = 0, 
\label{c7FE6}\\[1cm]
&& \ddot{c}_{ijmnrs} - 3 b_{i\;\fbox{$\scriptstyle mnr$}}b_{kkj\;\fbox{$\scriptstyle s$}} 
        - 3 b_{j\;\fbox{$\scriptstyle mnr$}}b_{ppi\;\fbox{$\scriptstyle s$}} 
        + 9 b_{ik\;\fbox{$\scriptstyle mn$}}b_{j\;\fbox{$\scriptstyle rs$}\;k} 
        + 9 b_{jk\;\fbox{$\scriptstyle mn$}}b_{i\;\fbox{$\scriptstyle rs$}\;k} 
        \nn\\
& &
 - 64 a_{i\;\fbox{$\scriptstyle m$}} a_{j\;\fbox{$\scriptstyle n$}} a_{\fbox{$\scriptstyle rs$}} 
        + 8 a_{i\;\fbox{$\scriptstyle m$}} c_{kkj\;\fbox{$\scriptstyle nrs$}} 
        + 8 a_{j\;\fbox{$\scriptstyle m$}} c_{kki\;\fbox{$\scriptstyle nrs$}} 
        - 4 a_{j\;\fbox{$\scriptstyle m$}} c_{ikk\;\fbox{$\scriptstyle nrs$}} 
        \nn\\ 
& &
- 4 a_{i\;\fbox{$\scriptstyle m$}} c_{jkk\;\fbox{$\scriptstyle nrs$}} 
        + 4 a_{ij} c_{kkmnrs}
        - \delta_{ij} \bigg( 3b_{\fbox{$\scriptstyle mnrs$}} b_{kkll} 
        + 3 b_{kk\;\fbox{$\scriptstyle mn$}}b_{ll\;\fbox{$\scriptstyle rs$}} 
        - 48 a_{\fbox{$\scriptstyle mn$}} a_{\fbox{$\scriptstyle rs$}} a_{kk} \bigg)
\nn \\ 
&& + 8\pi G \Bigg[ 4 \rho a_{\fbox{$\scriptstyle mn$}} V_{i\;\fbox{$\scriptstyle r$}}V_{j\;\fbox{$\scriptstyle s$}} 
        + \rho c_{ijmnrs} 
        + 2\rho V_{i\;\fbox{$\scriptstyle m$}}V_{j\;\fbox{$\scriptstyle n$}}V_{\fbox{$\scriptstyle rs$}}  \nn\\
&&
 \qquad + \delta_{ij} \left(\frac{1}{2}\rho c_{mnrskk}
                + 2\rho a_{\fbox{$\scriptstyle mn$}}V_{\fbox{$\scriptstyle r$}\;k}V_{k\;\fbox{$\scriptstyle s$}}
\right)\Bigg] = 0, 
\label{c7FE7} 
\end{eqnarray}
\vspace{0.5cm}
\noindent where we have introduced the notation
\begin{eqnarray*}
 b_{i\;\fbox{$\scriptstyle mnr$}}b_{ppj\;\fbox{$\scriptstyle s$}} & = & 
        \frac{1}{12} \bigg\{ b_{imnr} b_{ppjs} + b_{imrs} b_{ppjn} 
        + b_{imns} b_{ppjr} + b_{inmr} b_{ppjs} \nn\\
& & \quad + b_{inrs} b_{ppjm} + b_{inms} b_{ppjr} + b_{irmn} b_{ppjs} 
        + b_{irms} b_{ppjn} \nn\\
& & \quad + b_{irns} b_{ppjm} + b_{ismn} b_{ppjr} 
        + b_{ismr} b_{ppjn} + b_{isnr} b_{ppjm}\bigg\} 
\end{eqnarray*}
and similar, to indicate total symmetrisation in all indices that
are surrounded by a box.

The harmonic gauge conditions (\ref{c7hg2},\ref{c7hg1}) give
\begin{eqnarray}
\dot{a} & = & - \frac{1}{4}b_{kk}, \nn \\
\dot{a}_{ij} & = & - \frac{3}{4}b_{kkij}, \nn \\
\dot{b}_{ij} & = & 2 c_{ikkj}, \nn \\
\dot{b}_{ijkl} & = & 4 c_{immjkl}\;,\label{c7cond}
\end{eqnarray}
and the Bianchi identities become
\begin{eqnarray}
L(t) + c^{-2}(M(t) + N_{ij}(t)X_{ij}) = 0 \label{c7B1new}
\end{eqnarray}
and
\begin{eqnarray}
O_{ij}X_j + c^{-2}(Q_{ij}(t)X_j + S_{ijkl}(t)X_{jkl}) = 0,\label{c7B2}
\end{eqnarray}
where
\begin{eqnarray}
L(t) & = & \dot{\rho} + \rho V_{kk}, \nn \\
M(t) & = & - 4\dot{\rho} a + \frac{1}{2} \rho b_{kk} + p V_{kk} 
        - \frac{3}{8\pi G} a_{kk}b_{ll}, \nn \\
N_{ij}(t) & = &  \rho \left(V_{ik}V_{jk}V_{ll} + V_{ik}\dot{V}_{jk} 
        + V_{jk}\dot{V}_{ik} + V_{ik}V_{kl}V_{lj} + V_{jk}V_{kl}V_{li} 
        + \frac{3}{2}b_{kkij}\right) \nn \\
&& 
        + \frac{1}{2} \dot{\rho} \left(V_{ik}V_{kj} + V_{jk}V_{ki} 
                - 8 a_{ij} \right) 
        + \frac{1}{16\pi G} \bigg[\frac{7}{2} a_{ik}b_{kllj} 
        + \frac{7}{2} a_{jk}b_{klli} - 18a_{kk}b_{llij} \bigg]. \nn\\
O_{ij}(t) & = & \dot{V}_{ij} + V_{ik}V_{kj} + 2a_{ij}, \nn \\
Q_{ij}(t) & = & \frac{1}{\rho}\Bigg\{ 
        V_{ij} \left(4\dot{\rho}a +  4\rho \dot{a} - \frac{1}{2}\rho b_{kk} 
                - \rho a V_{kk} - \dot{p}\right) 
        + 24 \rho a a_{ij} - p\dot{V}_{ij}
        + \frac{1}{2}\dot{\rho}b_{ij} + \frac{1}{2}\rho\dot{b}_{ij}\nn \\
&& 
        \quad 
        - \left( p + \rho a \right) V_{ik}V_{kj} + \frac{p}{2} a_{ij}
        + 2 \rho a V_{ki}V_{kj} - \rho c_{ikkj}\Bigg\}\nn\\
& &
        + \frac{1}{16 \pi G \rho} \Bigg\{ 
        -6 V_{ij} a_{kk} b_{ll} + 6 \dot{b}_{kk} a_{ij} 
        + 6 b_{kk} \dot{a}_{ij} + 4 a_{ik} \dot{b}_{kj} 
        + 4 \dot{a}_{ik} b_{kj} -96 a a_{ik} a_{jk} \nn\\
& &
        \quad  
        - 48 a c_{klklij} + 3 b_{ij} b_{kkll}
        - 3 b_{ik} b_{llkj} - 6 b_{ik} b_{kllj}
        + 6 b_{kj} b_{ikll} - 6 b_{ki} b_{llkj}  \nn\\
& &
        \quad 
        + 9 b_{kk} b_{llij} - 9 b_{kj} b_{llik} - 6 b_{kl} b_{klij} 
        + 6 b_{kl} b_{lkij}  
        -12 a_{ij} c_{klkl} + 8 a_{ij} c_{kkll} - 8 a_{ik} c_{kllj} \nn\\
& &
        \quad + 4 a_{ik} c_{llkj} + 4 a_{jk} c_{llki} 
        + 8 a_{kl} c_{klij} - 8 a_{kj} c_{ilkl} - 4 a_{kk} c_{llij}  
        \Bigg\}\nn\\
S_{ijkl}(t) & = & \frac{\dot{\rho}}{\rho} \Bigg\{ 
        4 V_{i\;\fbox{$\scriptstyle j$}} a_{\fbox{$\scriptstyle kl$}} 
        + \frac{1}{2} b_{ijkl}
        - V_{i\;\fbox{$\scriptstyle j$}} V_{m\;\fbox{$\scriptstyle k$}}
                V_{m\;\fbox{$\scriptstyle l$}} \Bigg\}
        + \frac{1}{2} \dot{b}_{ijkl}
        + 4 V_{i\;\fbox{$\scriptstyle j$}} \dot{a}_{\fbox{$\scriptstyle kl$}}
        + 24 a_{i\;\fbox{$\scriptstyle j$}} a_{\fbox{$\scriptstyle kl$}} \nn\\
&&
        + \frac{3}{2} V_{i\;\fbox{$\scriptstyle j$}} 
                b_{mm\;\fbox{$\scriptstyle kl$}} 
        - 2 c_{immjkl} 
        - \dot{V}_{i\;\fbox{$\scriptstyle j$}} V_{m\;\fbox{$\scriptstyle k$}} 
                V_{m\;\fbox{$\scriptstyle l$}}
        - 4 a_{mi} V_{\fbox{$\scriptstyle jk$}} 
                V_{m\;\fbox{$\scriptstyle l$}} \nn\\
&&
        + 2 a_{i\;\fbox{$\scriptstyle j$}}
          \left[ V_{m\;\fbox{$\scriptstyle k$}} V_{m\;\fbox{$\scriptstyle l$}} 
        - V_{\fbox{$\scriptstyle k$}\;m} V_{m\;\fbox{$\scriptstyle l$}} 
        - V_{mm} V_{\fbox{$\scriptstyle kl$}} \right]
        - V_{im} V_{m\;\fbox{$\scriptstyle j$}} 
                V_{n\;\fbox{$\scriptstyle k$}} V_{n\;\fbox{$\scriptstyle l$}}
        \nn\\
&&
        + \frac{1}{16\pi G \rho} \Bigg\{ 
                - V_{i\;\fbox{$\scriptstyle j$}} 
                  \left[48 a_{m\;\fbox{$\scriptstyle k$}} 
                  \dot{a}_{m\;\fbox{$\scriptstyle l$}}
                + 18 a_{mm} b_{nn\;\fbox{$\scriptstyle kl$}} 
                - 24 a_{m\;\fbox{$\scriptstyle k$}} 
                  b_{mnn\;\fbox{$\scriptstyle l$}} \right]\nn\\
&&
        \qquad + 18 a_{i\;\fbox{$\scriptstyle j$}} 
                        \dot{b}_{mm\;\fbox{$\scriptstyle kl$}} 
                + 18 \dot{a}_{i\;\fbox{$\scriptstyle j$}} 
                        b_{mm\;\fbox{$\scriptstyle kl$}} 
                + 4 a_{im} \dot{b}_{m\;\fbox{$\scriptstyle jkl$}}
                + 4 \dot{a}_{im} b_{m\;\fbox{$\scriptstyle jkl$}} \nn\\
&&
        \qquad + 48 a_{i\;\fbox{$\scriptstyle j$}} 
                  a_{m\;\fbox{$\scriptstyle k$}} a_{m\;\fbox{$\scriptstyle l$}}
                + 64 a_{i\;\fbox{$\scriptstyle j$}} 
                  a_{mm} a_{\fbox{$\scriptstyle kl$}}  
                - 120 a_{i\;\fbox{$\scriptstyle j$}} 
                  c_{mnmn\;\fbox{$\scriptstyle kl$}}
                + 16 a_{mn} c_{mnijkl} \nn\\
&&
        \qquad - 48 a_{im} 
                   c_{\fbox{$\scriptstyle j$}\;nmn\;\fbox{$\scriptstyle kl$}}
                - 8 a_{mm} c_{nnijkl}
                + 56 a_{im} c_{nnmjkl}
                - 16 a_{im} c_{mmnjkl} \nn\\
&&
        \qquad + 24 a_{i\;\fbox{$\scriptstyle j$}} 
                  c_{mmnn\;\fbox{$\scriptstyle kl$}}
                - 9 b_{i\;\fbox{$\scriptstyle jk$}\;m} 
                  b_{nnm\;\fbox{$\scriptstyle l$}}
                + 3 b_{ijkl} b_{mmnn}
                + 6 b_{imnn} b_{mjkl} \nn\\
&&
        \qquad - 18 b_{mni\;\fbox{$\scriptstyle j$}} 
                  b_{mn\;\fbox{$\scriptstyle kl$}}
                - 18 b_{mnni} b_{mjkl} - 27 b_{mmni} b_{njkl}\nn\\
&&
       \qquad + 27 b_{mmi\;\fbox{$\scriptstyle j$}} 
                  b_{nn\;\fbox{$\scriptstyle kl$}}
                + 18 b_{mni\;\fbox{$\scriptstyle j$}} 
                  b_{nm\;\fbox{$\scriptstyle kl$}} \Bigg\}.
        \nn
\end{eqnarray}
Note that setting $L(t) = 0$, i.e. setting higher order
terms in $c^{-2}$ to zero, one recovers the continuity
equation of the Newtonian approximation. Similarly, $O(t) = 0$ yields the
Euler equation of the Newtonian approximation.

It can be shown that the time derivative of (\ref{c7FE1}) and the trace of
equation (\ref{c7FE3}) combined give
\begin{eqnarray}
L(t) + c^{-2} M(t) = 0, \label{c7B1a}
\end{eqnarray}
in accordance with the Bianchi identity (\ref{c7B1}). This essentially is the
continuity equation with $c^{-2}$ corrections. The time derivative of
equation (\ref{c7FE2}) and the trace of equation (\ref{c7FE4}) yield
\begin{eqnarray}
N_{ij}(t)X_{ij} = 0 \label{c7B1b}
\end{eqnarray}
which is also consistent with the Bianchi identity (\ref{c7B1}).

We can also recover the post-Newtonian
Euler equation by combining the time derivative of (\ref{c7FE3}) and the
diagonal elements of (\ref{c7FE6}) to obtain
\begin{eqnarray}
O_{ij}(t)X_j + c^{-2}Q_{ij}(t) X_j= 0.\label{c7B2a}
\end{eqnarray}

The final part
\begin{eqnarray}
S_{ijkl}X_{jkl}(t) = 0, \label{c7B2b}
\end{eqnarray}
can be obtained by combining the time derivative of (\ref{c7FE4}) with equation
(\ref{c7FE7}).

Thus we may completely define the anisotropic, inhomogeneous post-Newtonian
cosmology with  the set of equations;
(\ref{c7B1a}), (\ref{c7B2a}), (\ref{c7FE1}), (\ref{c7FE2}), (\ref{c7FE3}), (\ref{c7FE4}),
(\ref{c7FE7}), (\ref{c7FE6}, $m \ne n$) and
(\ref{c7FE5}).\footnote{%
Eq. (\ref{c7FE5}) is the only equation not to contribute to the Bianchi
identities.}
In the case of the isotropic, homogeneous post-Newtonian cosmology,
\cite{rain} it was shown that there existed relationships between the field
equations rendering many of them redundant. This is special to the case of
isotropy and due to the symmetries of  the potentials. We know very little
about the symmetries of the potentials in the more general situation of
anisotropy. The harmonic gauge conditions provide a little assistance. For
example:\\
Newtonian cosmology is contained within the equations (\ref{c7B1a}),
(\ref{c7B2a}), (\ref{c7FE1}), (\ref{c7FE2}), (\ref{c7FE3}), (\ref{c7FE4}), (\ref{c7FE7}),
(\ref{c7FE6}, $m \ne n$) and (\ref{c7FE5}), as the special case when the $c^{-2}$
terms go to zero. Equations (\ref{c7FE1}), (\ref{c7B1a}) and (\ref{c7B2a}) are then
the Newtonian theory with $c^{-2}$ corrections. The corrections to the theory
contain extra unknowns - the potentials $\zeta_i =  b_{ijkl} X_{jkl} +
b_{ij}X_j$ and $\phi_{ij} = c_{ijklmn}X_{klmn} + c_{ijkl}X_{kl} + c_{ij}$
which are determined by the equations  (\ref{c7FE2}), (\ref{c7FE3}), (\ref{c7FE4}),
(\ref{c7FE7}), (\ref{c7FE6}, $m \ne n$) and (\ref{c7FE5}).

We will now attempt to use the equations (\ref{c7B1a}), (\ref{c7B2a}),
(\ref{c7FE1}), (\ref{c7FE2}), (\ref{c7FE3}), (\ref{c7FE4}), (\ref{c7FE7}), (\ref{c7FE6},
$m \ne n$) and (\ref{c7FE5}), to solve for the unknowns; $a_{ij}(t)$, $a(t)$,
$b_{ijkl}(t)$, $b_{ij}(t)$, $c_{ijklmn}(t)$, $c_{ijkl}(t)$ and
$c_{ij}(t)$.

Let us start with equation (\ref{c7FE3}) which has trace
\begin{eqnarray}
3b_{lkkl} = 8 \pi G \rho V_{kk} + c^{-2}\left(\dot{c}_{lkkl} - 3a_{kk}b_{ll}
+ 8 \pi G \left[pV_{kk} - \frac{1}{2}\rho b_{kk}\right]\right). \label{c7E1}
\end{eqnarray}
With the aid of the harmonic gauge conditions (\ref{c7cond}), it can be shown
that this is just the time derivative of equation (\ref{c7FE1}). Thus this
equation is redundant. The symmetric part of (\ref{c7FE3}) is
\begin{eqnarray}
\lefteqn{\frac{3}{2} \left( b_{ikkj} + b_{jkki}\right) 
        + \frac{4}{3}\dot{a}_{kk}\delta_{ij} 
= 8\pi G \rho \sigma_{ij} }\label{c7E2}\\
& & 
+ c^{-2}\left\{
        \frac{1}{4} \left[\ddot{b}_{ij}+\ddot{b}_{ji}\right]
        + \frac{2}{3}\dddot{a} \delta_{ij} 
        + 12 \dot{a} a_{ij} - 4 \dot{a} a_{kk} \delta_{ij} 
        + 2\pi G\left[4 p \sigma_{ij} - \rho \left(b_{ij} + b_{ji}\right) 
        - \frac{8}{3}\rho\dot{a} \delta_{ij})\right]\right\}, \nn
\end{eqnarray}
where the harmonic gauge conditions and equation (\ref{c7FE1}) have been
used. The antisymmetric part is given by
\begin{eqnarray}
\lefteqn{8\pi G \rho \omega_{ij} + \frac{3}{2} \left( b_{jkki} - b_{ikkj}\right)}
\nn\\
& & 
        + c^{-2}\left\{ \frac{1}{4}\left[\ddot{b}_{ij} - \ddot{b}_{ji}\right] 
        - 2a_{ik}b_{kj} + 2a_{jk}b_{ki} + 2\pi G \left[4 p \omega_{ij}
        - \rho \left(b_{ij} - b_{ji}\right)\right] \right\} = 0,
\label{c7E3}
\end{eqnarray}
where once again the harmonic gauge conditions have been used for
simplification where ever possible. These equations can be solved once we
know $a_{ij}$ to obtain information about $b_{ij}$\footnote{$b_{kk}$ of
course being given by the harmonic gauge conditions.} and $b_{ijkl}$.

We now want to find $a_{ij}$. To do so we consider equations (\ref{c7FE2}),
(\ref{c7FE4}) and (\ref{c7FE7}). Equation (\ref{c7FE2}) has the trace
\begin{eqnarray} \ddot{a}_{kk} = 6a_{kl}a_{kl} 
- 4\pi G\rho \left\{4a_{kk} + \frac{1}{3} \theta^2 + 2 \sigma^2 + 2 \omega^2
\right\}. \label{c7E6}
\end{eqnarray}
providing us with an equation for $a_{kk}$. The traceless part is given by
\begin{eqnarray}
6 a_{ik}a_{jk} - \ddot{a}_{ij} 
        - 4\pi G \rho \left( 4 a_{ij} + \frac{1}{9} \theta^2 + 2 \sigma^2 
        + 2 \omega^2 + \frac{2}{3} \theta\sigma_{ij} + \sigma_{ki}\omega_{kj} 
        + \sigma_{kj} \omega_{ki} \right) = 0, \label{c7E7}
\end{eqnarray}
which yields $a_{ij}$. Summing over $k=l$  in (\ref{c7FE4}) gives an equation
which is just the time derivative of (\ref{c7E7}).

We still have information remaining in equation (\ref{c7FE4}), however. The
remainder of equation (\ref{c7FE4}) is
\begin{eqnarray}
\ddot{b}_{ijkl} = 
    18a_{i\;\fbox{$\scriptstyle j$}} b_{\fbox{$\scriptstyle kl$}\;mm} 
    - 12a_{m\;\fbox{$\scriptstyle j$}} b_{mi\;\fbox{$\scriptstyle kl$}} 
        + 4a_{im}b_{mjkl}
        - 8\pi G \rho \left(- b_{ijkl} 
        + 4 V_{i\;\fbox{$\scriptstyle j$}} V_{m\;\fbox{$\scriptstyle k$}} 
        V_{m\;\fbox{$\scriptstyle l$}} \right).\nn\\
\label{c7E9}
\end{eqnarray}

\noindent From this equation the traceless part of $b_{ijkl}$ can be
determined.

Consider equation (\ref{c7FE5}). The trace is given by
\begin{eqnarray}
\lefteqn{2 c_{kkll} - 16aa_{kk} + 48\pi G(2\rho a + p)} \nn\\
&& + c^{-2}\left(- \ddot{c}_{kk} -  b_{kl}b_{kl} + 4 a_{kk}c_{ll} -
8a_{kl}c_{kl} - 24\dot{a}^2 + 4\pi  G\left[5\rho c_{kk} -
3 ap\right]\right) = 0.\label{c7E4}
\end{eqnarray}
It can be shown that, when combined with (\ref{c7FE1}), this equation is just
(\ref{c7B2a}) again and thus is redundant. Now consider the symmetric  piece
\begin{eqnarray}
\lefteqn{2c_{ijkk} - \frac{2}{3}c_{mmkk} 
        - 16 a a_{ij} + \frac{16}{3} a a_{kk} \delta{ij} 
        + c^{-2}\Bigg\{-\ddot{c}_{ij} + \frac{1}{3}\ddot{c}_{kk}\delta_{ij} 
        + 2b_{ik}b_{jk} - \frac{2}{3}b_{kl}b_{kl}\delta_{ij}} \nn\\
& & 
        + 4a_{ij}c_{kk} - \frac{4}{3}a_{kk}c_{ll}\delta_{ij}
        - 4 a_{ik}c_{jk} - 4a_{jk}c_{ik} 
        + \frac{8}{3}a_{kl}c_{kl}\delta_{ij}
        + 8\pi G \rho\left[c_{ij} - \frac{1}{3} c_{kk}\delta_{ij}\right]
\Bigg\} = 0,\nn\\ \label{c7E5}
\end{eqnarray}
which can be used in equations (\ref{c7E2}) and (\ref{c7E3}) above to define
$c_{ij}$. There is no antisymmetric equation.

We have seen that equations (\ref{c7E6}) and (\ref{c7E7}) give $a_{ij}(t)$. The
harmonic gauge conditions can be used to determine $b_{kk}$, and equations
(\ref{c7E2}) and (\ref{c7E3}) give the trace-free part of $b_{ij}$ We may then
determine $b_{ijkl}$ in the following manner: Use the harmonic gauge
conditions to obtain $b_{kkij}$ (recall that there is symmetry in the last
three indices). Then use Eq.~(\ref{c7E9}) to obtain the remaining
$b_{ijkl}$'s. The $c_{ij}$'s come from equation (\ref{c7E5}). To solve for
$c_{ijkl}$ use the harmonic gauge conditions to find $c_{ikkj}$ (where
there is symmetry in the first two and second two indices) and
Eq.~(\ref{c7FE6}). Finally,  Equation (\ref{c7FE7}) then provides
$c_{ijklmn}$.

Thus, equations (\ref{c7FE2}), (\ref{c7FE3}), (\ref{c7FE4}), (\ref{c7FE5}) and
(\ref{c7FE7}) determine the unknowns  $a_{ij}$, $b_{ijkl}$, $b_{ij}$,
$c_{ijklmn}$, $c_{ijkl}$ and $c_{ij}$, and the set of equations (\ref{c7FE1}),
(\ref{c7B1a}), (\ref{c7B2a}), now contain only the eleven unknowns $\theta$,
$\omega_{ij}$, $\sigma_{ij}$, $\rho$ and $A(t)$. This means that the
equations can be solved for all the unknowns uniquely. 

Therefore the post-Newtonian approximation provides a well-posed, closed,
complete system. Thus, the post-Newtonian approximation produces a set of
cosmological equations which are more in the spirit of the Bianchi types of
general relativity.


\section{Newtonian Theory and the Post-Newtonian Approximation of Rotation-free
Anisotropic Homogeneous Cosmologies}

\subsection{Rotation-free Anisotropic Homogeneous Newtonian Cosmology}

We first like to examine the Newtonian approximation of cosmology with
vanishing rotation and diagonal shear, $\sigma_{ij} =
{\rm diag} (\sigma_{11}, \sigma_{22}, -\sigma_{11} - \sigma_{22})$. The
Newtonian approximation (Eqs. (\ref{c7poisson}) and (\ref{c7cont}) to
(\ref{c7sigma})) is then given by
\begin{eqnarray}
a_{ii} & = & 4\pi G \rho, \label{c7ent1} \\
\dot{\rho} + \rho \theta & = & 0, \label{c7ent2} \\
\dot{\theta} & = & 4\pi G \rho - \frac{1}{3} \theta^2 - 2 \sigma^2,
\label{c7theta} \\
\dot{\sigma}_{ij} & = & a_{ij} - \frac{2}{3} \theta \sigma_{ij} -
\sigma_{ik}\sigma_{kj} + \delta_{ij}(\frac{2}{3}\sigma^2  - \frac{1}{3}
a_{kk}). \label{c7sigij}
\end{eqnarray}

\noindent For $i\ne j$, equation (\ref{c7sigij}) yields
\begin{eqnarray}
a_{ij} = 0, \quad i \ne j.
\end{eqnarray}

\noindent The diagonal elements of the shear obey the differential equation
(\ref{c7sigij}, $i = j$, no summation over~$i$)
\begin{eqnarray}
\dot{\sigma}_{ii} = a_{ii} - \frac{2}{3} \theta \sigma_{ii} - \Sigma_{k}
\sigma_{ik} \sigma_{ki} + \frac{2}{3} \sigma^2 - \frac{1}{3} \Sigma_k
a_{kk}. \label{c7sigstuff}
\end{eqnarray}

\noindent Summing over $i$ just yields the definition of $\sigma ^2$
\begin{eqnarray}
2\sigma^2 = \sigma_{ij} \sigma_{ji} = 2 \left( \sigma_{11}^2 + \sigma_{22}^2
+ \sigma_{11}\sigma_{22}\right) . \label{c74445}
\end{eqnarray}
Thus,  (\ref{c7sigstuff}) provides at most two independent
equations. Introducing $\theta = 3 \frac{\dot{R}}{R}$ the set of equations
can be written as
\begin{eqnarray}
\frac{\ddot{R}}{R} & = & \frac{4}{3} \pi G \rho_0 R^{-3} - \frac{2}{3}
(\sigma_{11}^2 + \sigma_{22}^2 + \sigma_{11}\sigma_{22}), \label{c7*1R}
\\[0.5cm]
\rho & = & \rho_0 R^{-3}, \label{c7*2R} \\[0.5cm]
a_{11} & = & \dot{\sigma}_{11} + 2\frac{\dot{R}}{R} \sigma_{11} +
\frac{1}{3} \sigma_{11}^2 - \frac{2}{3}( \sigma_{22}^2 +
\sigma_{11}\sigma_{22}) + \frac{4}{3} \pi G \rho_0 R^{-3}, \label{c7*3} \\
a_{22} & = & \dot{\sigma}_{22} + 2\frac{\dot{R}}{R} \sigma_{22} +
\frac{1}{3} \sigma_{22} ^2 - \frac{2}{3}( \sigma_{11}^2 +
\sigma_{11}\sigma_{22}) + \frac{4}{3} \pi G \rho_0 R^{-3}, \label{c7*4}
\\[0.5cm]
a_{33} & = & 4\pi G \rho_0 R^{-3} - a_{11} - a_{22}, \label{c7*5}
\end{eqnarray}
where $\rho_0$ is a constant.

Consider the structure of these five equations: If we provide the functions
$\sigma_{11}$ and $\sigma_{22}$ for all time, then the theory has a well-posed
initial value problem for the variables $R(t)$, $\rho(t)$, $a_{11}(t)$,
$a_{22}(t)$ and
$a_{33}(t)$.  Equation (\ref{c7*1R}) is the Raychaudhuri equation (obtained
from (\ref{c7theta})), and its
solution provides $R(t)$. From Eq.~(\ref{c7*2R}) (which comes from
(\ref{c7ent2})) the density $\rho (t)$
can be extracted which, once an equation of state is provided, gives $p(t)$.
Equations (\ref{c7*3}) and (\ref{c7*4}) (from (\ref{c7sigij})) give $a_{11}$ and
$a_{22}$, and from
(\ref{c7*5}) (or (\ref{c7ent1})), which is the Poisson equation, $a_{33}$ can
be extracted.

Thus, for a homogeneous anisotropic cosmology in Newtonian theory, there is
an initial value problem provided we supply two functions for all time. We have
seen that the post-Newtonian theory is able to give as many equations as
unknowns. So there is no need to specify any functions of time. Actually,
there is no need to (arbitrarily) specify any of the unknown functions in
order to obtain an initial value problem. We will revisit the Newtonian
theory shortly, when we will consider the Bianchi I metric as a specific
example of a homogeneous anisotropic cosmology with vanishing rotation.

\subsection{The Homogeneous Bianchi I Universe}

\subsubsection{The Metric}

Now we would like to study Newtonian theory and the post-Newtonian
approximation as applied to a specific example of an anisotropic
homogeneous cosmology. To this end we consider the homogeneous
Bianchi I models - i.e. the Heckmann-Sch\"ucking solutions of general
relativity.%
\footnote{Not to be confused with the Heckmann-Sch\"ucking
cosmology of Newtonian theory.}
The metric of the general Bianchi I universe has the form \\

\vspace{0.5cm}

$ds^2 = -dx_0^2 + R_{ik}R_{jk}dx_idx_j$, \hspace{1cm} where
$x_0 = ct$ \hspace{1cm} with $R_{ij} = {\rm diag}(R_{11}, R_{22},
R_{33})$.

\vspace{0.5cm}

\noindent To proceed further, we need to write the Bianchi I metric into a
form off which
the potentials $\phi$, $\zeta_i$ and $\phi_{ij}$ can be read. To do so, we
consider the following transformation
\begin{eqnarray}
x_o & = & Tc + \tau c^{-1} + \tau' c^{-3}, \nn \\
x_i & = & R_{ij}^{-1}X_j + \chi_i c^{-2} + \chi_i' c^{-4} \label{c7ansatze2}
\end{eqnarray}
with
\begin{eqnarray}
\tau & = & A(t) + A_{ij}(t)X_{ij}, \nn \\
\tau' & = & B_{ij}(t)X_{ij} + B_{ijkl}(t)X_{ijkl}, \nn\\
\chi_{i} & = & C_{ij}(t)X_j + C_{ijkl}(t)X_{jkl}, \nn \\
\chi'_i & = & D_{ijkl}(t)X_{jkl} + D_{ijklmn}(t)X_{jklmn}, \nn
\end{eqnarray}
where $X_{ij} \equiv X_iX_j$ and similar for $X_{ijk}$ etc. Throughout the
remainder of the paper we assume: $\dot{A} \ll c^{2}$ to ensure
convergence of the expansion in $c^{-2}$. $ A(t)$, $A_{ij}(t)$, $B_{ij}(t)$,
$B_{ijkl}(t)$, $C_{ij}(t)$, $C_{ijkl}(t)$, $D_{ijkl}(t)$ and $D_{ijklmn}(t)$
are arbitrary functions of time. In these coordinates the metric reads
\begin{eqnarray}
\lefteqn{ds^2 = c^2 dT^2
\Bigg[-1 + c^{-2} \left(
        -2\dot{A} - 2\dot{A}_{ij} X_{ij} +
        \left(R^{-2}\dot{R}^2\right)_{ij} X_{ij}
   \right) } \nn \\
&&
\quad + c^{-4} \bigg\{
  \left(\dot{A} + \dot{A}_{ij} X_{ij}\right)^2 - 2\dot{B}_{ij}X_{ij}
  - 2 \dot{B}_{ijkl}X_{ijkl}
  - 2 \dot{R}_{im} \dot{C}_{mjkl} X_{ijkl}
  - 2 \dot{R}_{jl} \dot{C}_{il}X_{ij} \nn \\
&&
\quad - 2 \dot{A} \left(\dot{R}^2 R^{-2}\right)_{ij} X_{ij}
  - 2 \dot{A}_{kl} \left(\dot{R}^2 R^{-2}\right)_{ij} X_{ijkl}
\bigg\}
  + {\cal O} (c^{-6}) \Bigg] \nn \\
&&
+ cdTdX_i
\Bigg[ c^{-1} \left(
        -4A_{ij}X_j - 2 \left(\dot{R}R^{-1}\right)_{ij} X_j
  \right) \nn\\
&& 
\quad + c^{-3} \bigg( - 8\dot{A}A_{ij}X_j - 8A_{ij}\dot{A}_{kl}X_{jkl}
  - 4B_{ij} X_j - B_{ijkl}X_{jkl} 
  + 4 A_{il} \left(\dot{R}^2 R^{-2}\right)_{jk} X_{jkl}
  \nn\\
&&
\qquad + 2\dot{R}_{jl} C_{il}X_j + 2R_{il}\dot{C}_{lj}
  - 6\dot{R}_{jm} C_{mikl}X_{jkl}
  + 2R_{im}\dot{C}_{mjkl}X_{jkl} \nn\\
&&
\qquad + 2\dot{A}\left(\dot{R} R\right)_{ij} X_j
        + 2\dot{A}_{kl}\left(\dot{R} R\right)_{ij} X_{jkl}
  \bigg) + {\cal O}(c^{_5})
\Bigg] \nn \\
&& + dX_i dX_j \Bigg[
  \delta_{ij} + c^{-2} \left(
        -4A_{ik}A_{jl}X_{kl} + 4A_{jk}\left(\dot{R} R \right)_{il} X_{kl}
        + 2 R_{ik}C_{jk} + 6R_{im}C_{mjkl}X_{kl}
  \right)\nn \\
&&
\quad + c^{-4} \bigg(
  - 8\dot{A}A_{ik}A_{jl}X_{kl} - 8\dot{A}_{mn}A_{ik}A_{jl}X_{klmn}
  - 8B_{jk}A_{il}X_{kl} - 16 B_{iklm}A_{jn}X_{klmn} 
\nn \\
&&
\qquad + 6R_{im}D_{mjkl}X_{kl} + 10R_{ip}D_{pjklmn}X_{klmn} 
   + R^2_{mn} C_{im}C_{jn} + 9R_{ip}R_{jq}C_{pikl}C_{qjmn}X_{klmn} 
\nn \\
&&
\qquad + 3 R^2_{mn}C_{im}C_{njkl}X_{kl} + 3R_{mk}R_{kn}C_{jn}C_{mikl}X_{kl}
  + 4 \left(\dot{R}^2R^{-2}\right)_{kl} A_{im}A_{jn}X_{klmn} \nn \\
&&
\qquad + 4 A_{jl}\dot{R}_{km} C_{mi}X_{kl} 
  + 12A_{jn}\dot{R}_{kp}C_{pilm}X_{klmn}
  + 4A_{jm}R_{il}\dot{C}_{kl}X_{km} + 4A_{jn}R_{pi}\dot{C}_{pklm}X_{klmn} \nn
\\
&&
\qquad + 4\dot{A}\left(\dot{R}R^{-1}\right)_{ik}A_{jl}X_{kl}
  + 4 \left(\dot{R}R^{-1}\right)_{ik} \dot{A}_{mn}A_{jl}X_{klmn}
  + 4 \left(\dot{R}R^{-1}\right)_{ik} B_{jl}X_{kl} \nn\\
&&
\qquad  + 8 \left(\dot{R}R^{-1}\right)_{ik} B_{jlmn}X_{klmn}
\bigg) + {\cal O}(c^{-6}) \Bigg].
\label{c7*metric}
\end{eqnarray}

It  can be seen from (\ref{c7pnform}) that there are no terms of order $c^{-1}$,
thus
\begin{eqnarray}
A_{ij} = - \frac{1}{2} \left(\dot{R} R^{-1}\right)_{ij}. \nn
\end{eqnarray}

Also, from (\ref{c7pnform}) we see that the $c^{-2}$ terms in $g_{00}$ and
$g_{ij}$ must be identical. Hence,
\begin{eqnarray}
C_{ij} & = & \dot{A}R^{-1}_{ij}, \nn \\
C_{ijkl} & = & \frac{1}{4} \left[
\left(\dot{R} R^{-2} \right)_{il} \left(\dot{R} R^{-1}\right)_{jk} 
+ \left(\dot{R} R^{-2} \right)_{ik} \left(\dot{R} R^{-1}\right)_{jl} \right]
+ \frac{1}{6} R^{-1}_{ij} \left(\ddot{R} R^{-1}\right)_{kl}. \nn
\end{eqnarray}

We may now read off the potentials $\phi$, $\zeta_i$ and $\phi_{ij}$ from the
metric (\ref{c7*metric}). With the help of the harmonic gauge conditions
(\ref{c7hg2}) and
(\ref{c7hg1}) we are able to write down the functions $a(t)$,
$a_{ij} (t)$ etc. which appear in the ans\"atze (\ref{c7ansatze}):
\begin{eqnarray}
\lefteqn{a(t) = \dot{A},} \nn \\
\lefteqn{a_{ij}(t) = - \frac{1}{2} \left(\ddot{R} R^{-1}\right)_{ij}\;,} \nn \\
\lefteqn{b_{ij}(t) = - 4 B_{ij} + 6 \dot{A} \left(\dot{R} R^{-1} \right)_{ij}
   - 2 \ddot{A} \delta_{ij} \;,} \nn \\
\lefteqn{b_{ijkl}(t) = - 8 B_{ijkl} + \frac{1}{36} \left(
          3 \delta_{i\;\fbox{j}} \left(\dddot{R} R^{-1} \right)_{\fbox{kl}}
        + \delta_{\fbox{jk}} \left(\dddot{R} R^{-1}\right)_{i\;\fbox{l}}
\right)} \nn \\
&&
- 5 \left(\dot{R} R^{-1}\right)_{i\;\fbox{j}} 
        \left(\dot{R}^2 R^{-2}\right)_{\fbox{kl}}
- \frac{8}{3} \left(\dot{R} R^{-1}\right)_{i\;\fbox{j}} 
        \left(\ddot{R} R^{-1}\right)_{\fbox{kl}} \;, \nn \\
\lefteqn{c_{ij}(t) = 2 \dot{A}^2 \delta_{ij},} \nn \\
\lefteqn{c_{ijkl}(t) = - \delta_{ij}\left(\dot{B}_{kl}+\dot{B}_{lk}\right)
        - 3 \delta_{ij} \dot{A} \left(\dot{R}^2 R^{-2} \right)_{kl}
        - 2 \delta_{ij} \dot{A} \left(\ddot{R} R^{-1} \right)_{kl} 
        + 2 \delta_{ij} \ddot{A} \left(\dot{R}R^{-1} \right)_{kl}} \nn\\
&&
        + \frac{1}{2} \delta_{ik} \left(\dot{R}R^{-1}\right)_{jl} 
        + \frac{1}{2} \delta_{jk} \left(\dot{R}R^{-1}\right)_{il} 
        + \frac{1}{2} \delta_{il} \left(\dot{R}R^{-1}\right)_{jk}
        + \frac{1}{2} \delta_{jl} \left(\dot{R}R^{-1}\right)_{ik} \nn\\
&&
        + 3R_{im}D_{mjkl} + 3R_{jm}D_{mikl} 
        + 2B_{ik} \left(\dot{R}R^{-1}\right)_{jl} 
        + 2B_{il} \left(\dot{R}R^{-1}\right)_{jk} 
        + 2B_{jk} \left(\dot{R}R^{-1}\right)_{il} \nn \\
&&
        + 2B_{jl} \left(\dot{R}R^{-1}\right)_{ik} 
        - \frac{11}{4} \dot{A} \left(\dot{R}R^{-1}\right)_{ik} 
                \left(\dot{R}R^{-1}\right)_{jl} 
        - \frac{11}{4} \dot{A} \left(\dot{R}R^{-1}\right)_{il}
                \left(\dot{R}R^{-1}\right)_{jk} \nn \\
&&
        - \frac{3}{4}\dot{A} \left(\dot{R}R^{-1}\right)_{ik}
                \left(\ddot{R}R^{-1}\right)_{jl}
        - \frac{3}{4}\dot{A} \left(\dot{R}R^{-1}\right)_{il}
                \left(\ddot{R}R^{-1}\right)_{jk} \nn\\
&&
        - \frac{3}{4}\dot{A} \left(\dot{R}R^{-1}\right)_{jk}
                \left(\ddot{R}R^{-1}\right)_{il} 
        - \frac{3}{4}\dot{A} \left(\dot{R}R^{-1}\right)_{jl}
                \left(\ddot{R}R^{-1}\right)_{ik}\;, \nn \\
\lefteqn{c_{ijklmn}(t) =- 2 \delta_{ij} \dot{B}_{\fbox{klmn}}
        + \frac{1}{4}\left(\dot{R}^2R^{-2}\right)_{\fbox{kl}}
                \left(\dot{R}^2R^{-2}\right)_{\fbox{mn}} \delta_{ij}}\nn\\
&&
        - \frac{1}{6}\delta_{ij} \left( \ddot{R}R^{-1} \right)_{\fbox{kl}}
                \left(\dot{R}^2R^{-2}\right)_{\fbox{mn}}
        + \frac{1}{2}\delta_{ij} \left( \ddot{R}R^{-1} \right)_{\fbox{kl}}
                \left(\ddot{R}R^{-1} \right)_{\fbox{mn}} \nn\\
&&
        + 2\delta_{ij} \left(\dot{R}R^{-1} \right)_{\fbox{kl}}
                \left(\dot{R}^3R^{-3}\right)_{\fbox{mn}}
        - \frac{2}{3}  \left(\dot{R}R^{-1} \right)_{\fbox{kl}}
                \left(\ddot{R}\dot{R}R^{-2}\right)_{\fbox{mn}} \nn\\
&&
        - \frac{1}{3} \delta_{ij} \left(\dot{R}R^{-1}\right)_{\fbox{kl}}
                \left( \dddot{R}R^{-1} \right)_{\fbox{mn}}
        + 5 R_{io} D_{ojklmn} + 5R_{jo}D_{oiklmn}\nn\\
&&
        + 8 B_{i\;\fbox{klm}} \left(\dot{R}R^{-1} \right)_{j\;\fbox{n}} 
        + 8 B_{j\;\fbox{klm}} \left(\dot{R}R^{-1} \right)_{i\;\fbox{n}} \nn\\
&&
        - \frac{1}{6}\left(\dot{R}R^{-1}\right)_{i\;\fbox{k}} 
                \left(\dot{R}R^{-1}\right)_{j\;\fbox{l}}
                \left(\ddot{R}R^{-1}\right)_{\fbox{mn}} \nn \\
&&  
        + \frac{25}{4}\left(\dot{R}R^{-1}\right)_{i\;\fbox{k}} 
                \left(\dot{R}R^{-1}\right)_{j\;\fbox{l}}
                \left(\dot{R}^2R^{-2}\right)_{\fbox{mn}} \nn\\
&&
        + \frac{1}{6} \delta_{i\;\fbox{k}} 
                \left(\dot{R}R^{-1}\right)_{\fbox{jl}}
                \left(\ddot{R}\dot{R}R^{-2}\right)_{\fbox{mn}}
        - \frac{1}{6} \delta_{i\;\fbox{k}} 
                \left(\dot{R}R^{-1}\right)_{\fbox{jl}}
                \left(\dddot{R}R^{-1}\right)_{\fbox{mn}} \nn\\
&&
        + \frac{1}{6} \delta_{j\;\fbox{k}} 
                \left(\dot{R}R^{-1}\right)_{\fbox{il}}
                \left(\ddot{R}\dot{R}R^{-2}\right)_{\fbox{mn}}
        - \frac{1}{6} \delta_{j\;\fbox{k}} 
                \left(\dot{R}R^{-1}\right)_{\fbox{il}}
                \left(\dddot{R}R^{-1}\right)_{\fbox{mn}} \nn\\
&&  
        + \left(\dot{R}^2R^{-2} \right)_{i\;\fbox{k}} 
                \left(\dot{R}R^{-1} \right)_{j\;\fbox{l}} 
                \left(\dot{R}R^{-1}\right)_{\fbox{mn}} 
        + \left(\dot{R}^2R^{-2} \right)_{j\;\fbox{k}} 
                \left(\dot{R}R^{-1} \right)_{i\;\fbox{l}} 
                \left(\dot{R}R^{-1}\right)_{\fbox{mn}} 
\nn \\
&&
        - \frac{1}{2} \left(\ddot{R}R^{-1} \right)_{i\;\fbox{k}}
                \left(\dot{R}R^{-1}\right)_{j\;\fbox{l}}
                \left(\dot{R}R^{-2}\right)_{\fbox{mn}}
        - \frac{1}{2} \left(\ddot{R}R^{-1} \right)_{j\;\fbox{k}}
                \left(\dot{R}R^{-1}\right)_{i\;\fbox{l}}
                \left(\dot{R}R^{-2}\right)_{\fbox{mn}}. \nn\\
\label{c7*potent}
\end{eqnarray}
The Heckmann-Sch\"ucking solution has $R_{ij} = {\rm diag} \left( R_{11},
R_{22}, R_{33}\right)$ such that
\begin{eqnarray}
R_{11} & = & (t - t_0)^{p_1} (t - t_1)^{q_1}, \label{c7*7a} \\
R_{22} & = & (t - t_0)^{p_2} (t - t_1)^{q_2}, \label{c7*7b} \\
R_{33} & = & (t - t_0)^{p_3} (t - t_1)^{q_3} \label{c7*7c}
\end{eqnarray}
with
\begin{eqnarray}
\Sigma_i p_i =\Sigma_i p_i^2  = 1, \label{c7*8}
\end{eqnarray}
where
\begin{eqnarray}
q_i = \frac{2}{3} - p_i \label{c7*9}
\end{eqnarray}
such that
\begin{eqnarray}
\Sigma_i q_i =\Sigma_i q_i^2  = 1. \label{c7*10}
\end{eqnarray}
These conditions can be fulfilled by the following
parameterisation for $p_1$ to $p_3$ and $q_1$ to $q_3$:
\begin{eqnarray}
 p_1 & := & u, \nn\\
 p_2 & := & \frac{1-u}{2} + \frac{1}{2} \sqrt{1+2u-3u^2},
\nn\\
 p_3 & := & \frac{1-u}{2} - \frac{1}{2} \sqrt{1+2u-3u^2},
\nn\\
 q_1 & := & \frac{2}{3} - u, \nn\\
 q_2 & := & \frac{1+3u}{6} - \frac{1}{2} \sqrt{1+2u-3u^2},
\nn\\
 q_3 & := & \frac{1+3u}{6} + \frac{1}{2} \sqrt{1+2u-3u^2}. \label{c7upara}
\end{eqnarray}

\subsubsection{The Heckmann-Sch\"ucking Solution of General Relativity and
Newtonian Theory}

The Bianchi I model is rotation-free and so the Newtonian theory is just
given by equations (\ref{c7*1R}) to (\ref{c7*5}). From (\ref{c7*potent}) we see
that $a_{ij} = - \frac{1}{2}\frac{\ddot{R}_{ik}}{R_{jk}}$.
Define $R(t)$ such that
\begin{eqnarray}
\theta = 3 \frac{\dot{R}}{R} =
\frac{(R_{11}R_{22}R_{33})^.}{R_{11}R_{22}R_{33}}
= \frac{2t-t_0-t_1}{(t-t_0)(t-t_1)} . \label{c7deftheta}
\end{eqnarray}
The parameters to be determined are $u$, $t_0$, $t_1$ and $\rho_0$, and the
variables are $\rho$, $\sigma_{11}$, $\sigma_{22}$.

First consider equations (\ref{c7*3}) and (\ref{c7*4}) which using the
parametrisation (\ref{c7upara}) become:
\begin{eqnarray}
\lefteqn{\dot{\sigma_{11}} + \frac{2}{3} \sigma_{11}
\frac{2t-t_0-t_1}{(t-t_0)(t-t_1)} + \frac{1}{3}\sigma_{11}^2 -
\frac{2}{3}(\sigma_{22}^2 + \sigma_{11}\sigma_{22}) + \frac{4}{3} \pi G \rho
\frac{1}{(t-t_0)(t-t_1)} =} \nn\\
& & \hspace*{40ex}
- \frac{u(u-1)}{2(t-t_0)^2} + \frac{(2-3u)(1+3u)}{18(t-t_1)^2} 
        - \frac{u(2-3u)}{3(t-t_0)(t-t_1)} \nn
\end{eqnarray}
and
\begin{eqnarray}
\lefteqn{\dot{\sigma_{22}} + \frac{2}{3} \sigma_{22}
\frac{2t-t_0-t_1}{(t-t_0)(t-t_1)} + \frac{1}{3}\sigma_{22}^2 -
\frac{2}{3}(\sigma_{11}^2 + \sigma_{11}\sigma_{22}) + \frac{4}{3} \pi G \rho
\frac{1}{(t-t_0)(t-t_1)} =} \nn\\
& & 
\hspace*{35ex}
- \frac{\left(1-u+\sqrt{1+2u-3u^2}\right)\left(1+u-\sqrt{1+2u-3u^2}\right)}
        {8(t-t_0)^2} \nn\\
& &
\hspace*{35ex}
+ \frac{\left(1+3u-3\sqrt{1+2u-3u^2}\right)\left(5-u+3\sqrt{1+2u-3u^2}\right)}
        {72(t-t_1)^2} \nn\\
& &
\hspace*{35ex}
- \frac{\left(1-u+\sqrt{1+2u-3u^2}\right)\left(1+3u-3\sqrt{1+2u-3u^2}\right)}
        {12(t-t_0)(t-t_1)}. \nn
\end{eqnarray}

\vspace{0.5cm}

\noindent Once the parameters $u$, $t_0$ and $t_1$ are known we
are able to solve these two equations for $\sigma_{11}$ and
$\sigma_{22}$.\footnote{The condition 
$\sigma_{11} + \sigma_{22} + \sigma_{33} = 0$ determines
$\sigma_{33}$.} To proceed further, consider Eq.~(\ref{c7*1R})
which is a Raychaudhuri-like equation and may be written as
\begin{eqnarray}
\frac{4}{3} \pi \rho_0\frac{1}{(t-t_0)(t-t_1)} - \frac{2}{3}(\sigma_{11}^2 +
\sigma_{22}^2 + \sigma_{11}\sigma_{22}) =
\frac{1}{3((t-t_0)(t-t_1))^{\frac{5}{3}}} - \frac{2}{9}
\frac{2t-t_0-t_1}{((t-t_0)(t-t_1))^{\frac{8}{3}}}.\nn
\end{eqnarray}
If $\sigma_{11}$ and $\sigma_{22}$ are known then this equation
relates the two constants $t_0$ and $t_1$. The Poisson equation (\ref{c7*5})
simply defines $\rho_0$,
\begin{eqnarray}
\rho_0 = \frac{1}{12\pi G}.\nn
\end{eqnarray}
Finally, we can write down (\ref{c7*2R}) as
\begin{eqnarray}
\rho = \rho_0 \frac{1}{(t-t_0)(t-t_1)}, \nn
\end{eqnarray}
which determines $\rho(t)$. Therefore the shear variables, $\sigma_{11}$ and
$\sigma_{22}$, and the density $\rho$ are now expressed in terms of two
parameters with the initial condition $\rho (t=0)$ specifying one of them. 

Thus, the Newtonian theory in the homogeneous, anisotropic
rotation-free case can be solved, provided we arbitrarily fix the 
parameter~$u$ for all time. However, varying the equation of state
will have no effect on the solutions for the density, since the pressure in
no way enters the dynamics, as was the case in \cite{rain} with the FRW
cosmology.

\subsubsection{The Post-Newtonian Approximation of the Homogeneous Bianchi I
Universe}

The post-Newtonian theory for the homogeneous anisotropic rotation-free case
is given by equations (\ref{c7FE2}), (\ref{c7FE3}), (\ref{c7FE4}),
(\ref{c7FE7}), (\ref{c7FE6}, $m \ne n$) and (\ref{c7FE5}) which may be solved for
the unknowns $a_{ij}$, $b_{ij}$, $b_{ijkl}$, $c_{ij}$, $c_{ijkl}$ and 
$c_{ijklmn}$ as we have shown in section 2. The set of equations (\ref{c7FE1}),
(\ref{c7B1a}), and (\ref{c7B2a}) then determine the eight unknowns $\theta$,
$\sigma_{ij}$, $\rho$ and $A(t)$ (or $a(t)$). Vanishing rotation imposes
extra symmetry on the equations. Therefore many become redundant and there
are a great many simplifications in what does remain. Also we can use the
potentials (\ref{c7*potent}) for further simplification. Clearly $a_{ij} = 0$
for $i \ne j$, since $a_{ij} = -\frac{1}{2}\frac{\ddot{R}_{ik}}{R_{jk}}$,
with $R_{ij} = diag(R_{11}, R_{22},R_{33})$. 

Using the parametrisation (\ref{c7upara}) the Poisson-like equation (\ref{c7FE1})
becomes
\begin{eqnarray}
 \frac{1}{(t-t_0)(t-t_1)} & = & 6\pi G \rho + c^{-2} \left\{
24\pi G \rho
+ \frac{3}{2} \dddot{A} \right\}, \label{c7poiB1}
\end{eqnarray}
which can be used to determine the potential $A(t)$.

The Friedmann-like equation (\ref{c7B1a}), as usual, provides $\rho(t)$:
\begin{eqnarray}
 \dot{\rho} + \rho \frac{2t-t_0-t_1}{(t-t_0)(t-t_1)} +
c^{-2} \Bigg\{-4\dot{\rho}\dot{A}
  -2\rho\ddot{A} + p \frac{2t-t_0-t_1}{(t-t_0)(t-t_1)}
+\frac{\ddot{A}}{2\pi G} \frac{1}{(t-t_0)(t-t_1)} \Bigg\} =
0.
\end{eqnarray}

The Bianchi identity (\ref{c7B2a}) can be decomposed into a trace part,
symmetric trace-less part and an antisymmetric part. The trace yields
the Raychaudhuri-like equation
\begin{eqnarray}
\lefteqn{\frac{6(t-t_0)(t-t_1) - 2(2t-t_0-t_1)^2}{(t-t_0)^2(t-t_1)^2}
        + 4\pi G \left(\rho + 3p c^{-2}\right) - \sigma^2} \nn\\
& & 
- c^-2 \Bigg\{ \frac{\dot{A}}{9} 
        \left[\frac{124(2t-t_0-t_1)^2}{(t-t_0)^2(t-t_1)^2}
        - \frac{14 }{(t-t_0)(t-t_1)}\right] \nn\\
& &
\quad +\frac{22}{3} \ddot{A} \frac{2t-t_0-t_1}{(t-t_0)(t-t_1)}
        + \frac{123}{8} \dddot{A} + \sigma^2 \left[ -\frac{5}{3} \dot{A} 
        + \frac{3}{8} c_{kkll} (t-t_0)(t-t_1) \right]\nn\\
& &
\quad + c_{kkll} \left( \frac{63}{16} 
        - \frac{1}{2}\frac{(2t-t_0-t_1)^2}{(t-t_0)(t-t_1)}\right)
        + \dot{c}_{kkll} (2t-t_0-t_1) 
        - \frac{9}{4} b_kl \left(b_{klmm} + b_{lkmm}\right) \nn\\
& &
\quad - 96 \dot{A} \left[
        \left( \frac{u(3u-2)}{3(t-t_0)(t-t_1)}
        - \frac{u(u-1)}{2(t-t_1)^2}
        + \frac{(2-3u)(3u+1)}{18(t-t_0)^2} \right)^2 \right.\nn\\
& &
\quad\quad 
   + \left\{\frac{\left(u-1-\sqrt{1+2u-3u^2}\right)
        \left(-u-1+\sqrt{1+2u-3u^2} \right)}{4(t-t_0)(t-t_1)}\right.\nn\\
& &
\quad\hspace{26ex} 
        + \frac{\left(1-u + \sqrt{1+2u-3u^2} \right)
          \left(u+1-\sqrt{1+2u-3u^2} \right)}{8(t-t_1)^2}\nn\\
& &
\quad\hspace{26ex}
        - \left.\frac{\left(3u+1-3\sqrt{1+2u-3u^2} \right)
          \left(3u-5 -3\sqrt{1+2u-3u^2} \right)}{72(t-t_0)^2} \right\}^2 \nn\\
& &
\quad\quad 
   + \left\{\frac{\left(u-1+\sqrt{1+2u-3u^2}\right)
        \left(3u+1+3\sqrt{1+2u-3u^2} \right)}{12(t-t_0)(t-t_1)}\right.\nn\\
& &
\quad\hspace{26ex} 
        + \frac{\left(1-u-\sqrt{1+2u-3u^2} \right)
          \left(u+1-\sqrt{1+2u-3u^2} \right)}{8(t-t_1)^2}\nn\\
& &
\quad\hspace{26ex} 
        - \left.\left.\frac{\left(3u+1+3\sqrt{1+2u-3u^2}\right)
          \left(3u-5+3\sqrt{1+2u-3u^2} \right)}{72(t-t_0)^2} \right\}^2 
        \right]\nn\\
& &
\quad + \frac{1}{24} \dot{b}_{11} \left[-6u \left(3u-2\right) 
        + \frac{9u(u-1)(t-t_0)}{t-t_1}
        + \frac{(3u-2)(3u+1)(t-t_1)}{t-t_0} \right] \nn\\
& &
\quad + \frac{1}{96} \dot{b}_{22} \Bigg[ 6\left(1-u+\sqrt{1+2u-3u^2}\right)
          \left(1+3u-3\sqrt{1+2u-3u^2}\right)\nn\\
& &
\quad\quad
        - \frac{9\left(1-u+\sqrt{1+2u-3u^2}\right)
          \left(1+u-\sqrt{1+2u-3u^2}\right)(t-t_0)}{t-t_1}\nn\\
& &
\quad\quad
        + \left.\frac{\left(1+3u-3\sqrt{1+2u-3u^2}\right)
          \left(-5+3u-3\sqrt{1+2u-3u^2}\right)(t-t_1)}{t-t_0}\right]\nn\\
& &
\quad + \frac{1}{96} \dot{b}_{33} \Bigg[ 6\left(1-u-\sqrt{1+2u-3u^2} \right)
          \left(1+3u+3\sqrt{1+2u-3u^2}\right)\nn\\
& &
\quad\quad
        - \frac{9\left(1-u-\sqrt{1+2u-3u^2}\right)
          \left(1+u+\sqrt{1+2u-3u^2}\right)(t-t_0)}{t-t_1}\nn\\
& &
\quad\quad
        + \left.\frac{\left(1+3u+3\sqrt{1+2u-3u^2}\right)
          \left(-5+3u+3\sqrt{1+2u-3u^2}\right)(t-t_1)}{t-t_0}\right]\nn\\
& &
\quad + \frac{1}{6} b_{11} \left[\frac{3u(2-3u)(2t-t_0-t_1)}{(t-t_0)(t-t_1)}
        + \frac{9u(u-1)(t-t_0)}{(t-t_1)^2} 
        - \frac{(2-3u)(3u+1)(t-t_1)}{(t-t_0)^2} \right]\nn\\
& &
\quad + \frac{1}{48} b_{22} \left[\frac{6\left(-1+u - \sqrt{1+2u-3u^2}\right)
          \left(1+3u-3\sqrt{1+2u-3u^2}\right)(2t-t_0-t_1)}{(t-t_0)(t-t_1)}
\right.\nn\\
& &
\quad\quad
        + \frac{12\left(-1-u+\sqrt{1+2u-3u^2}\right)
          \left(1+u-\sqrt{1+2u-3u^2}\right) (t-t_0)}{(t-t_1)^2}\nn\\
& &
\quad\quad
        - \left.\frac{\left(1+3u-3\sqrt{1+2u-3u^2}\right)
          \left(-5+u-3\sqrt{1+2u-3u^2}\right)( t-t_1)}{(t-t_0)^2} 
        \right]\nn\\
& &
\quad + \frac{1}{48} b_{33} \left[ \frac{6\left(-1+u+\sqrt{1+2u-3u^2}\right)
          \left(1+3u+3\sqrt{1+2u-3u^2}\right)(2t-t_0-t_1)}{(t-t_0)(t-t_1)}
\right.\nn\\
& &
\quad\quad
        + \frac{12\left(1-u-\sqrt{1+2u-3u^2}\right)
          \left(1+u+\sqrt{1+2u-3u^2}\right) (t-t_0)}{(t-t_1)^2}\nn\\
& &
\quad\quad
        - \left.\left.\frac{\left(1+3u+3\sqrt{1+2u-3u^2}\right)
          \left(-5+u+3\sqrt{1+2u-3u^2}\right)( t-t_1)}{(t-t_0)^2}\right]
 \right\} =0,
\end{eqnarray}
which may be used to determine $u$.

The traceless symmetric part is given by
\begin{eqnarray}
 \lefteqn{\dot{\sigma}_{ij} + \frac{2}{3} \sigma_{ij}
\frac{2t-t_0-t_1}{(t-t_0)(t-t_1)}
  + \sigma^2_{ij} - \frac{2}{3} \sigma^2 \delta_{ij} + 2
a_{ij}
 -\frac{2}{9(t-t_0)(t-t_1)} \delta_{ij} = } \nn\\
 & &
 c^{-2} \Bigg\{
 \frac{1}{12}\sigma_{ij} \left[106\ddot{A} 
        - 56\dot{A} \frac{2t-t_0-t_1}{(t-t_0)(t-t_1)}
        +3(t-t_0)(t-t_1)\dot{c}_{kkll}
        +3\left(2t-t_0-t_1\right){c}_{kkll} \right] \nn\\
& &
\quad
        +\frac{70}{3}\dot{A}a_{ij}
        +\frac{1}{8}(t-t_0)(t-t_1)c_{kkll}
                \left(23a_{ij}+2\dot{\sigma}_{ij}\right)
        +\frac{4}{3}\dot{A}\dot{\sigma}_{ij}\nn\\
& &
\quad
        +\frac{2\left[\dot{A}+9\ddot{A}(2t-t_0-t_1)\right]}
                {27(t-t_0)(t-t_1)}\delta_{ij} 
        -\frac{1}{8}\left(\dot{b}_{ij}+\dot{b}_{ji}\right)
        -\frac{23}{72}c_{kkll}\delta_{ij} \nn\\
& &
\quad
   + (t-t_0)(t-t_1) \bigg[
        -3\dddot{A} a_{ij} +9\ddot{A} \dot{a}_{ij}
        +\frac{9}{2}\left(\dot{a}_{ik}b_{kj}+\dot{a}_{Jk}b_{ki}\right)
        -\frac{7}{2} \dot{a}_{kl}b_{kl}\delta_{ij}\nn\\
& &
\quad\quad
        -36a a_{ik}a_{jk} + 12 a a_{kl}a_{kl} \delta_{ij}
        +\frac{3}{4} \left(\dot{a}_{jk} b_{ik}+\dot{a}_{ik} b_{jk}\right)\nn\\
& &
\quad\quad
        -\frac{3}{4} \left(b_{ik}b_{kjll}+b_{jk}b_{kill}
           +3b_{kl}b_{klij}+3b_{kl}b_{lkij}
           -b_{kl}b_{klmm}\delta_{ij}-b_{kl}b_{lkmm}\delta_{ij}\right)\nn\\
& &
\quad\quad
        +\frac{3}{2}\left(a_{ik}c_{lljk}+a_{jk}c_{llik}\right)
        -a_{km}c_{llkm}\delta_{ij}+3a_{kl}c_{klij}
 \bigg]\Bigg\},
\end{eqnarray}
and may be used to find $\sigma_{11}$, $\sigma_{22}$ and $\sigma_{33} =
-\sigma_{11}-\sigma_{22}$.

The antisymmetric piece of Eq.~(\ref{c7B2a}) is
\begin{eqnarray}
\lefteqn{4(\dot{a}_{ik}{kj}-\dot{a}_{jk}b_{ki})+2(\dot{a}_{jk}b_{ik}
-\dot{a}_{ik}b_{jk}-a_{jk}\dot{b}_{ik}+a_{ik}\dot{b}_{jk})}\nn\\
& &
\quad\hspace*{30ex} = 
3(b_{ik}b_{kjll}-b_{jk}b_{kill}-b_{kj}b_{ikll}+b_{ki}b_{jkll}).
\label{c7antiB1}
\end{eqnarray}
These are three equations. However, with the aid of the harmonic gauge
conditions it can be shown that only two of the equations are
independent. These can be used to determine $t_0$ and $t_1$.

Thus, the Heckmann-Sch\"ucking solutions of general relativity for
the post-Newtonian approximation are defined by the eight variables
$\rho$, $R_{11}$, $R_{22}$, $R_{33}$,  $\sigma_{11}$, $\sigma_{22}$
$\sigma_{33}$, $A(t)$,\footnote{And, $p(t)$ which is defined through a
barotropic equation of state.} which are completely  determined by the
equations (\ref{c7poiB1}) to (\ref{c7antiB1}).

A second improvement on
the Newtonian theory is that here pressure enters into the dynamics through
the Raychaudhuri equation. This means that varying the equation of state
varies the solutions for the density $\rho$, the cosmic scale factor $R(t)$
and the shear $\sigma (t)$. As in the FRW cosmology \cite{rain} the function
of time $A(t)$ is what incorporates the pressure into the system.

\section{Newtonian Theory and the Post-Newtonian Approximation of Shear-Free
Anisotropic Homogeneous Cosmologies}

\subsection{The Newtonian Shear-Free Anisotropic Homogeneous Universe}

The  Heckmann-Sch\"ucking solution of section 2.1 is the Newtonian
approximation of an
anisotropic homogeneous cosmology. The cosmology is given by equations
(\ref{c7poisson}), (\ref{c7cont}) and (\ref{c7teta}) to (\ref{c7sigma}) with
vanishing shear. The equations reduce to
\begin{eqnarray}
a_{ii} = 4\pi 
G \rho, \\
\rho = \rho_0 R^{-3}, \label{c7*1} \\
\omega_i = \omega_{i0} R^{-2}, \label{c7**1} \\
\frac{\ddot{R}}{R} = - \frac{4 \pi G \rho_0}{3R^3} + \frac{2\omega_0^2}{R^4}.
\label{c7Ray*}
\end{eqnarray}
Recall that with the five shear functions arbitrarily set to zero we
have a determined system. The final equation (\ref{c7Ray*}) may be integrated
out to give the Heckmann-Sch\"ucking solution
\begin{eqnarray}
\dot{R}^2 = \frac{8\pi G \rho_0}{3R} - \frac{2\omega_0^2}{3R^2} -
\frac{\epsilon}{\tau_0^2}, \label{c7Ray**}
\end{eqnarray}
where $\epsilon = \pm\; 1\; {\rm or} \; 0$, $\tau_o$ an arbitrary constant.
When $\omega_{i0} = 0$ we have a shear-free and rotation-free cosmology which
becomes isotropic. This is just FRW, which is the most general possible
solution for homogeneity and isotropy. In this case equation (\ref{c7Ray**})
can be identified with the Raychaudhuri equation of the FRW cosmology.

A theorem of Ellis \cite{ellisthrm} states that in the case of
shear-free dust either the expansion or the rotation must vanish. Setting
$\theta = 0$, we obtain the following solution for $R(t)$

\begin{eqnarray}
R = \frac{2 \omega_o^2}{8\pi G \rho_0}.
\label{c7result}
\end{eqnarray}

\noindent Hence $R$ is a constant, and as long as $\omega_0$ is non-zero,
there is no possibility of there being a singularity for $R(t)$. The solution
will have minimum and maximum values for $R(t)$ but may never be
zero, \cite{szek-rank}.

Next we explore the analogous case in the post-Newtonian approximation
and see how far it comes in overcoming this difficulty.

\subsection{The Post-Newtonian Approximation of an Anisotropic Homogeneous
Shear-free Universe}

The post-Newtonian approximation for the shear-free case is given by the field
equations (\ref{c7FE2}), (\ref{c7FE3}), (\ref{c7FE4}),
(\ref{c7FE7}), (\ref{c7FE6}, $m \ne n$) and
(\ref{c7FE5}), along with a Poisson-like equation (\ref{c7FE1}) and the Bianchi
identities (\ref{c7B1a}),
(\ref{c7B2a}), with $\sigma_{ij} = 0$. The equations (\ref{c7FE1}), (\ref{c7B1a}) and
(\ref{c7B2a}) are eleven equations, and the variables are reduced to $\rho$
(which may be determined from (\ref{c7B1a})), $\theta$ (which may be determined
from the trace of (\ref{c7B2a})), $\omega_{ij}$ (which may be determined from the
antisymmetric part of (\ref{c7B2a})) and $a(t)$ (which may be determined from
the Poisson-like equation (\ref{c7FE1})). Unlike in the rotation-free case where
the $a_{ij}$ can be determined from the additional field equations here we will
use what remains of the symmetric part of (\ref{c7B2a}) to determine the
$a_{ij}$ for $i \ne j$. The unknowns $b_{ij}$, $b_{ijkl}$, $c_{ij}$,
$c_{ijkl}$ and $c_{ijklmn}$ that
appear in the $c^{-2}$ corrections to (\ref{c7FE1}), (\ref{c7B1a}) and
(\ref{c7B2a}) are determined from the additional field equations (\ref{c7FE2}),
(\ref{c7FE3}), (\ref{c7FE4}),
(\ref{c7FE5}), (\ref{c7FE6}, $m \ne n$) and
(\ref{c7FE7}).

The reduction in unknowns must be accompanied by redundancies
otherwise the system will
be overdetermined. We saw earlier that equations (\ref{c7E6}) and (\ref{c7E7})
yield $a_{ij}$. Somewhat tedious calculations show that the time derivative
of the symmetric part
of (\ref{c7FE3}) and (\ref{c7FE6}) give the symmetric part of (\ref{c7B2a}) in
this special case of vanishing shear. Thus, we may ignore the symmetric part
of (\ref{c7B2a}).

The Poisson-like
equation (\ref{c7FE1}) is
\begin{eqnarray}
a_{kk} = 2\pi G \rho + c^{-2}(- \frac{1}{4}c_{klkl} + 8 \pi G \rho a),
\label{c7p}
\end{eqnarray}
which defines $a(t)$. Defining $\theta = 3\frac{\dot{R(t)}}{R(t)}$ we may
write the continuity equation (\ref{c7B1a}) which allows us to solve
for $\rho$ as\footnote{Which is the same as in the rotation-free case.}
\begin{eqnarray}
\left(
        \frac{\left(
                \left( R_{11}R_{22}R_{33}
                \right)^{\frac{1}{3}}
        \right)^.}{
        \left( R_{11}R_{22}R_{33}
        \right)^{\frac{1}{3}}}
\right)^2
= \frac{8 \pi G}{3} \rho + \gamma c^{-2},
\label{c7f}
\end{eqnarray}
where $\gamma$ is a solution of the differential equation
\begin{eqnarray}
&& \dot{\gamma} + 4 \gamma
\frac{\left(
        \left(
                \left(R_{11}R_{22}R_{33}
                \right)^{\frac{1}{3}}
        \right)^{..}
\right)}{
\left( R_{11}R_{22}R_{33}
\right)^{\frac{1}{3}}} 
+ \ddot{A}
\frac{\left(
        \left( R_{11}R_{22}R_{33}
        \right)^{\frac{1}{3}}
\right)^{..}}{
\left( R_{11}R_{22}R_{33}
\right)^{\frac{1}{3}}}\nn \\
&& 
+ 8\dot{a}\left(
        \frac{\left(
                \left( R_{11}R_{22}R_{33}
                \right)^{\frac{1}{3}}
        \right)^{..}
        \left(
                \left( R_{11}R_{22}R_{33}
                \right)^{\frac{1}{3}}
        \right)^.}{
        \left( R_{11}R_{22}R_{33}
        \right)^{\frac{2}{3}}}
        - \frac{\left(
                \left( R_{11}R_{22}R_{33}
                \right)^{\frac{1}{3}}
        \right)^{...}}{
        \left( R_{11}R_{22}R_{33}
        \right)^{\frac{1}{3}}}
\right)
= 0.
\end{eqnarray}
The antisymmetric part of (\ref{c7B2a}) becomes
\begin{eqnarray}
&& \dot{\omega}_{ij} + \frac{2}{3}\theta \omega_{ij} + \frac{1}{\rho} c^{-2}
\left( \omega_{ij}
        \left( -2\rho \dot{a} - 2 \dot{\rho} a + 2 \rho a \theta
                + \frac{1}{2}\rho b_{kk}
        \right) - \frac{1}{2} \rho
        \left( c_{ikjk} - c_{jkik}
        \right) 
\right. \nn \\
&&                + 2 \rho a \dot{\omega}_{ij} - \frac{4}{3} \rho a \theta \omega_{ij}
        + \frac{1}{16 \pi G}
        \left[ - \frac{8}{3}\theta \omega_{ij} - \frac{4}{3} \omega_{ij}
                \left( \dddot{a} + 4\dot{a}a_{kk} + 4a\dot{a}_{kk}
                \right)
        \right. \nn \\
&&  \qquad            - \frac{4}{3}
                \left( \dot{\omega}_{ij} + \frac{2}{3}\theta \omega_{ij}
                        + \frac{1}{2} \omega_{ik}\omega_{kj}
                        - \frac{1}{2} \omega_{jk}\omega_{ki}
                \right)
                \left( \ddot{a} + 4aa_{kk}
                \right) - 24 \omega_{ij}a_{kk}\dot{a} - 4 \dot{a}_{kj}b_{ki}\nn \\
&& \qquad       + 4\dot{a}_{ki}b_{kj} - 4 a_{kj}\dot{b}_{ki}
                + 4a_{ki}\dot{b}_{kj} - 15 b_{ik}b_{llkj}
                + 15 b_{jk}b_{llki} - 12a_{ik}c_{llkj} \nn \\
&& \qquad       + 12 a_{jk}c_{llki} - 4 a_{ik}c_{kljl}
                + 4a_{jk}c_{klil} - 4a_{jk}c_{ilkl} + 4a_{ik}c_{jlkl}
                + 8 a_{lk}c_{likj} \nn \\
&& \qquad   
\left.
        \left.
          - 8a_{lk}c_{ljki} - 8 a_{lk}c_{iklj} + 8a_{lk}c_{jkil}
        \right]
\right) = 0,
\label{c7dw}
\end{eqnarray}
which may be solved for $\omega_{ij}$. Finally, the trace of (\ref{c7B2a}) gives a
Raychaudhuri-like equation which is
\begin{eqnarray}
3 \frac{\ddot{R}}{R} & = & 4 \pi G
\left( \rho + 3pc^{-2}
\right) + 2\omega^2 + c^{-2}
\left(-114 a
        \left( \frac{\dot{R}}{R}
        \right)^2 + 20 \dot{a} \frac{\dot{R}}{R} + 8 a a_{kk}
        - \frac{95}{16} \ddot{a} 
\right.
\nn \\
&&
        - \frac{51}{2}c_{klkl} + \frac{10}{3}a\omega^2 + \frac{1}{4\pi G \rho}
\left[ 
                -6\ddot{a}
                \left( \frac{\dot{R}}{R}
                \right)^2 - \ddot{a} \frac{\ddot{R}}{R}
                - \frac{1}{3}\dddot{a} - \frac{2}{3}\ddot{a}\omega^2 
\right.\nn\\
&& \qquad
\left.
        \left.- \frac{1}{4}a a_{kl}a_{kl} + a_{kl}
                \left( 2c_{kmlm} - 8c_{mmkl} - 6c_{mlkm}
                \right)
                + \dot{b}_{kl}
        \right]
\right).
\label{c7r}
\end{eqnarray}

Thus, the case of an anisotropic homogeneous shear-free cosmology is
essentially defined by a Poisson-like equation (\ref{c7p}), a Friedmann-like
equation (\ref{c7f}), a Raychaudhuri-like equation (\ref{c7r}), and has rotation
given by (\ref{c7dw}). Its variables are $\rho (t)$, $R(t)$, $\omega_{ij} (t)$,
$a_{ij}(t)$ and $a(t)$.

In contrast to Newtonian theory, in post-Newtonian theory the Raychaudhuri
equation has the pressure entering the dynamics and so gives
rise to a variety of possible solutions for $\rho(t)$ and $R(t)$. Before the
only solution was that of dust. It is through the function
$a(t)$, that the pressure enters into the dynamics. Without
such a term the $c^{-2}$-corrections would vanish and we would be left with a
system where pressure is not dynamic. There is however, a more serious
reason for keeping these higher order
terms and it relates to the Ellis theorem: The Raychaudhuri-like equation
(\ref{c7r}), allows for solutions with $R(t) =
0$. Due to the pressure becoming dynamic, with a variation in equation of
state the
solutions for the density vary. Thus equations (\ref{c7*1}) and (\ref{c7**1})
no longer hold. Solutions different to dust are possible, and the rotation
is no longer just a constant. Hence the Raychaudhuri equation (\ref{c7r}) will
no longer yield the result (\ref{c7result}). 

\section{Conclusion}

In this paper we have derived the post-Newtonian approximation for
anisotropic homogeneous cosmologies. In contrast to the
Newtonian approximation the equations are well-posed. The cosmological
equations of the post-Newtonian approximation are much more in the spirit of
the Bianchi types of general relativity.

We considered a particular example of the Bianchi identities - the Bianchi I
cosmology. In the Newtonian theory assuming anisotropy and homogeneity
leads to the well known
Heckmann-Sch\"ucking solutions. Howsoever, even with homogeneity, there are
still not enough
equations to solve for all the unknowns and some need to be supplied for all
time. The post-Newtonian theory is able to overcome this problem and also
allows the pressure to enter into the dynamics of the theory. Therefore the
full set of possible solutions for $R(t)$ as outlined in \cite{rain}, can be
reproduced.

The result of pressure entering into the dynamics allows for more than just a
matter dominated universe. In this way more general solutions are obtained
than in the Newtonian case. In the shear-free case with rotation the
Raychaudhuri equation gives rise to a singularity. In this way the Ellis
theorem does not lead to contradictions, and there is hope that the solutions
of the post-Newtonian approximation may all have general relativistic
counterparts. In the very least we no longer have solutions to which there
can be no general relativistic analogues.

It seems reasonable then to replace the Newtonian theory with the
post-Newtonian approximation when considering homogeneous cosmologies,
whether they be isotropic or not.

\end{document}